\documentclass{emulateapj}
\usepackage{psfig}

\newcommand{\kms}{\mbox{km s$^{-1}~$}} 
\newcommand{\kmse}{\mbox{km s$^{-1}$}} 
 
\newcommand{\msun}{M$_{\odot}~$} 
\newcommand{\msune}{M$_{\odot}$} 
\newcommand{\vlsr}{$V_{\rm LSR}~$}
\newcommand{\vlsre}{$V_{\rm LSR}$}
\newcommand{\vlgsr}{$V_{\rm LGSR}~$}

\newcommand{\vgsre}{$V_{\rm GSR}$}
\newcommand{\lms}{$L_{\rm MS}~$}
\newcommand{\lmse}{$L_{\rm MS}$}
\newcommand{\bms}{$B_{\rm MS}~$}
\newcommand{\bmse}{$B_{\rm MS}$}
\newcommand{\dgr}{$^{\circ}~$}

\newcommand{\h}{$^{\rm h}$}
\newcommand{\m}{$^{\rm m}$}

\newcommand{\nhi}{$N_{\rm H \small{I}}$ }
\newcommand{\nhie}{$N_{\rm H \small{I}}$}

\newcommand{\hi}{H{\footnotesize I} }

\begin{document}

\title{The 200\degr--long Magellanic Stream System}

\shorttitle{The 200\degr--long Magellanic Stream System}
\shortauthors{NIDEVER ET AL.}

\author{David L. Nidever\altaffilmark{1},
Steven R. Majewski\altaffilmark{1}, 
W. Butler Burton\altaffilmark{2,3},
and Lou Nigra\altaffilmark{4}
}

\altaffiltext{1}{Dept. of Astronomy, University of Virginia,
Charlottesville, VA, 22904-4325 (dnidever, srm4n@virginia.edu)}

\altaffiltext{2}{Sterrewacht Leiden, PO Box 9513, 2300 RA Leiden,
The Netherlands}

\altaffiltext{3}{National Radio Astronomy Observatory, 520 Edgemont Road,
Charlottesville, Virginia 22903, USA (bburton@nrao.edu)}

\altaffiltext{4}{Dept. of Astronomy, University of Wisconsin,
475 North Charter St., Madison, WI 53706-1582, USA (nigra@astro.wisc.edu)}

\begin{abstract}
We establish that the Magellanic Stream (MS) is some 40\dgr longer than previously known with certainty
and that the entire MS and Leading Arm (LA) system is thus at least 200\dgr long.
With the Green Bank Telescope, we conducted a $\sim$200 deg$^2$, 21--cm survey at the
tip of the MS to substantiate the continuity of the MS between the Hulsbosch \& Wakker data and the
 MS--like emission reported by Braun \& Thilker.  Our survey, in combination with the Arecibo survey
by Stanimirovi{\'c} et al., shows that the MS gas is continuous in this region and that the MS is 
at least $\sim$140\dgr long.  The MS--tip is composed of a multitude of forks and filaments.  We identify
a new filament on the eastern side of the MS that significantly deviates from the equator of the MS coordinate
system for more than $\sim$45\degr.
Additionally, we find a previously unknown velocity inflection in the MS--tip near MS longitude
\lmse$\approx-120$\dgr at which the velocity reaches a minimum and then starts to increase.
We find that five compact high velocity clouds cataloged by de Heij et al.\ as well as Wright's Cloud
are plausibly associated with the MS because they match the MS in position and velocity.
The mass of the newly-confirmed $\sim$40\dgr extension of the MS--tip is $\sim$$2\times10^7$ \msun $(d/120~{\rm kpc})^2$
(including Wright's Cloud increases this by $\sim$50\%)
and increases the total mass of the MS by $\sim$4\%.
However, projected model distances of the MS at the tip are generally quite large and,
if true, indicate that the mass of the extension might be as large as $\sim$$10^8$ \msune.
From our combined map of the entire MS, we find that
the total column density (integrated transverse to the MS) drops markedly along the MS and follows an
exponential decline with \lms of \nhie=$5.9\times10^{21} \exp(L_{\rm{MS}}$/19.3\degr) cm$^{-2}$.
Under the assumption that the observed sinusoidal velocity pattern of the LMC filament of the MS is due to the
origin of the MS from a rotating LMC,
we estimate that the age of the $\sim$140\degr--long MS is $\sim$2.5 Gyr.  This coincides
with bursts of star formation in the Magellanic Clouds and a possible close encounter of these two
galaxies with each other that could have triggered the formation of the MS.
These newly observed characteristics of the MS offer additional constraints for MS simulations.
In the Appendix we describe a previously little discussed problem with a 
standing wave pattern in GBT \hi data and detail a method for removing it.
\end{abstract}


\section{Introduction}
\label{sec:intro}

More than three decades ago, evidence of a galactic \hi stream around the Milky Way (MW) was discovered as
an $\sim$100\degr--long \hi tail emanating from the Magellanic Clouds (MCs) and
named the Magellanic Stream \citep[MS;][]{WW72,Math74}.
The Leading Arm (LA), a counterpart to the MS that
stretches in front of the MCs, has also been known for many decades and suspected to be from the MCs \citep{Math74}, but
was not definitively linked to the MCs until the studies by \citet{Put98} and \citet{Lu98}.
The complex and double--filamentary structure of the MS was first clearly shown by \citet{Wakker01}, based on the data of 
\citet[][hereafter HW88]{HW88}, and later shown in more detail in the higher-resolution HIPASS data \citep{Barnes01} by
\citet[][hereafter P03]{Put03}.

\begin{figure}[t]
\includegraphics[angle=0,scale=0.345]{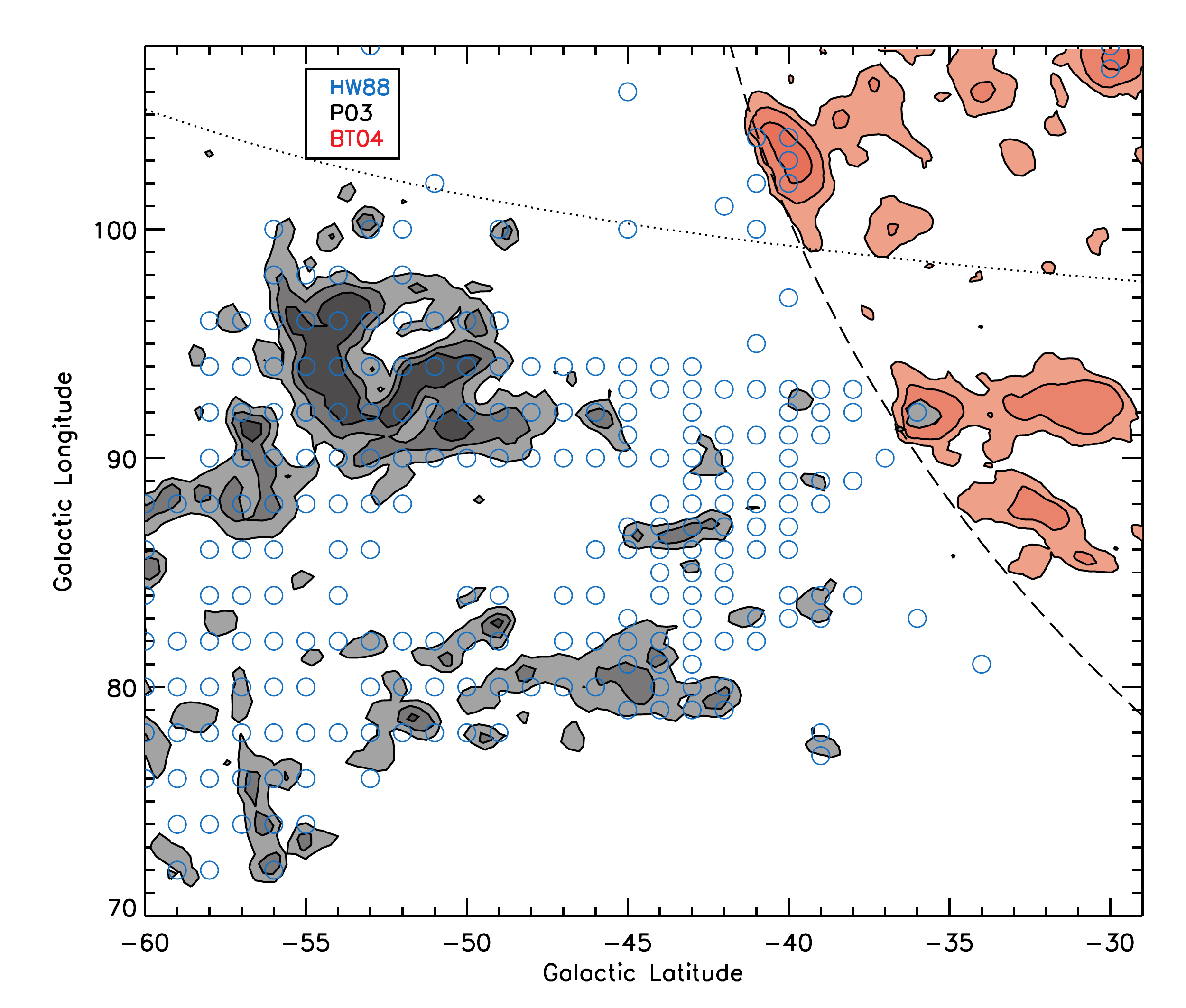}
\caption{\hi detections of the Magellanic Stream in the region of our survey by previous authors: open blue circles
indicate MS detections by HW88; the grayscale image shows \nhi from P03 with contour levels
of  0.2, 0.5, 1$\times10^{19}$ atoms cm$^2$; and the pink image shows \nhi from BT04 with contour levels at
0.2, 1, 5$\times10^{18}$ atoms cm$^2$.  The dotted line shows the edge of the P03 data; the dashed line, the edge
of the BT04 data.}
\label{fig_prevsurveys}
\end{figure}

Recently, \citet[][hereafter BT04]{BT04} conducted a deep and wide-field \hi survey of the M31 region with the Westerbork
Synthesis Radio Telescope (WSRT).
They serendipitously discovered some faint emission in the western portion of their
survey that appeared to be consistent with a spatial and kinematical extension of the MS, but at much lower column densities
(Fig.\ \ref{fig_prevsurveys}).
This suggested that the MS was plausibly longer than previously recognized, although a $\sim$10\dgr gap existed between
the MS as revealed by P03 and the faint emission discovered by BT04 (Fig.\ \ref{fig_prevsurveys} as well as Fig.\ 5 of BT04).
While the HW88 data show the MS continuing to the edge of the BT04 survey, these data lacked sufficient resolution and coverage
to be certain (Fig.\ \ref{fig_prevsurveys}).  A better map of this region would help to clarify the continuity of the MS in
this region.

Recent simulations by \citet{Besla07}, motivated by new proper motion measurements for the MCs \citep{Kalli06a,Kalli06b,Piatek08},
conclude that the MCs might be on their first passage around the MW.
The new orbit predictions suggest that the MCs have spent most of their lives isolated from the MW,
a scenario that has major consequences for
the viability of the two primary MS mechanisms that have traditionally been postulated for the formation of the MS
-- ram pressure stripping \citep[e.g.,][]{Meurer85,Sofue94,MD94,Mastro05} and tidal stripping \citep[e.g.,][]{MF80,GN96,YN03,C06}.
To remove the MS gas effectively, both of these mechanisms require the MCs to be relatively close to the MW for a prolonged
encounter time.  If, as suggested by \citet{Besla07}, the orbits of the MCs have kept them more isolated in the past,
then it becomes difficult for either ram pressure or tidal forces alone to create the MS.
The finding of an even longer MS would exacerbate this problem for the two standard mechanisms since a greater length
would require gas to be pulled out of the MCs at even larger MW--distances where the density of a ram pressure medium and
tidal forces are even lower.

However, \citet{Nidever08} put forward a third MS formation mechanism after (1) tracing one filament of the MS as well as
the LA back to their origin in the southeast \hi overdensity (SEHO) in the LMC, and (2) illustrating that the SEHO contains
large gaseous outflows from supergiant shells.
The ``blowout hypothesis'' postulates that supergiant shells in the dense SEHO blow out 
gas from the LMC to large radii where it is easier for ram pressure and/or tidal forces to fully strip the gas and 
disperse it to create the MS and LA.
Because blowout does the hard work of overcoming a significant portion of the gravitational hold of the LMC on the gas, this
scenario should be an effective solution to the large--distance dilemma raised by the \citet{Besla07} first-passage scenario.
In a proposed alternative scenario, \citet{Mastro09} was able to produce a 120\degr--long MS with ram pressure (and tidal forces)
using a high-velocity, unbound LMC orbit, but whether this new model can account for an even longer MS remains unclear.
A better understanding of the true length of the MS would place important, more specific constraints on
these various MS formation scenarios.

In this paper we address the question of whether the BT04 emission really is an extension of the
MS and if the MS is actually much longer than previously widely accepted.  In pursuit of this goal
we conducted a 200 deg$^2$ Green Bank Telescope \citep[GBT;][]{Lockman98} 21--cm survey in the region
between the tip of the ``classical'' MS and the features found in the BT04 survey to explore whether there is
continuity of the MS across this region.
Our GBT data, in combination with the Arecibo survey of \citet[][hereafter S08]{Stani08}, demonstrate that the MS
{\em is both spatially and kinematically continuous} across this region and therefore that the MS is
at least 140\dgr long ($\sim$40\dgr longer than
previously verified), and probably even longer.  This additional extension of the MS makes the entire MS system (including the
LA) at least 200\dgr long.
In addition, we combine all available \hi data of the MS--tip into one
datacube to investigate the structural characteristics of the MS in this region.  The MS--tip is composed of many thin
filaments, as previously noted by S08, but we identify a new filament in the eastern portion of the MS that markedly deviates
from the equator of the MS coordinate system for some 45\degr.
Additionally, we find that there is a previously unknown velocity inflection at the MS--tip where
the MS radial velocities reach a minimum and then begins to increase.
The mass of the newly mapped portion of the MS--tip is $\sim$$2\times10^7$ \msun $(d/120~{\rm kpc})^2$ which increases the estimated mass
of the entire MS by $\sim$4\%.

This paper is organized as follows:
A brief description of our GBT survey and data reduction is given in \S \ref{sec:obsred}, while more details on the data
reduction and some difficulties we encountered in dealing with GBT 21--cm spectral data
are presented in the Appendix.  In \S \ref{sec:combined}
we combine the various extant \hi datacubes of the MS tip and analyze the structural characteristics
of the extended MS.  Figures and descriptions of the entire 200\dgr MS+LA system are presented in \S \ref{sec:allms}.
A discussion of our findings is given in \S \ref{sec:discussion}, and a summary of our primary conclusions is given in
\S \ref{sec:summary}.

\section{GBT Survey}
\label{sec:obsred}

\subsection{Observations}
\label{subsec:gbtobs}

\begin{figure*}[ht!]
\begin{center}$
\begin{array}{cc}
\epsscale{1.00}
\includegraphics[angle=0,scale=0.45]{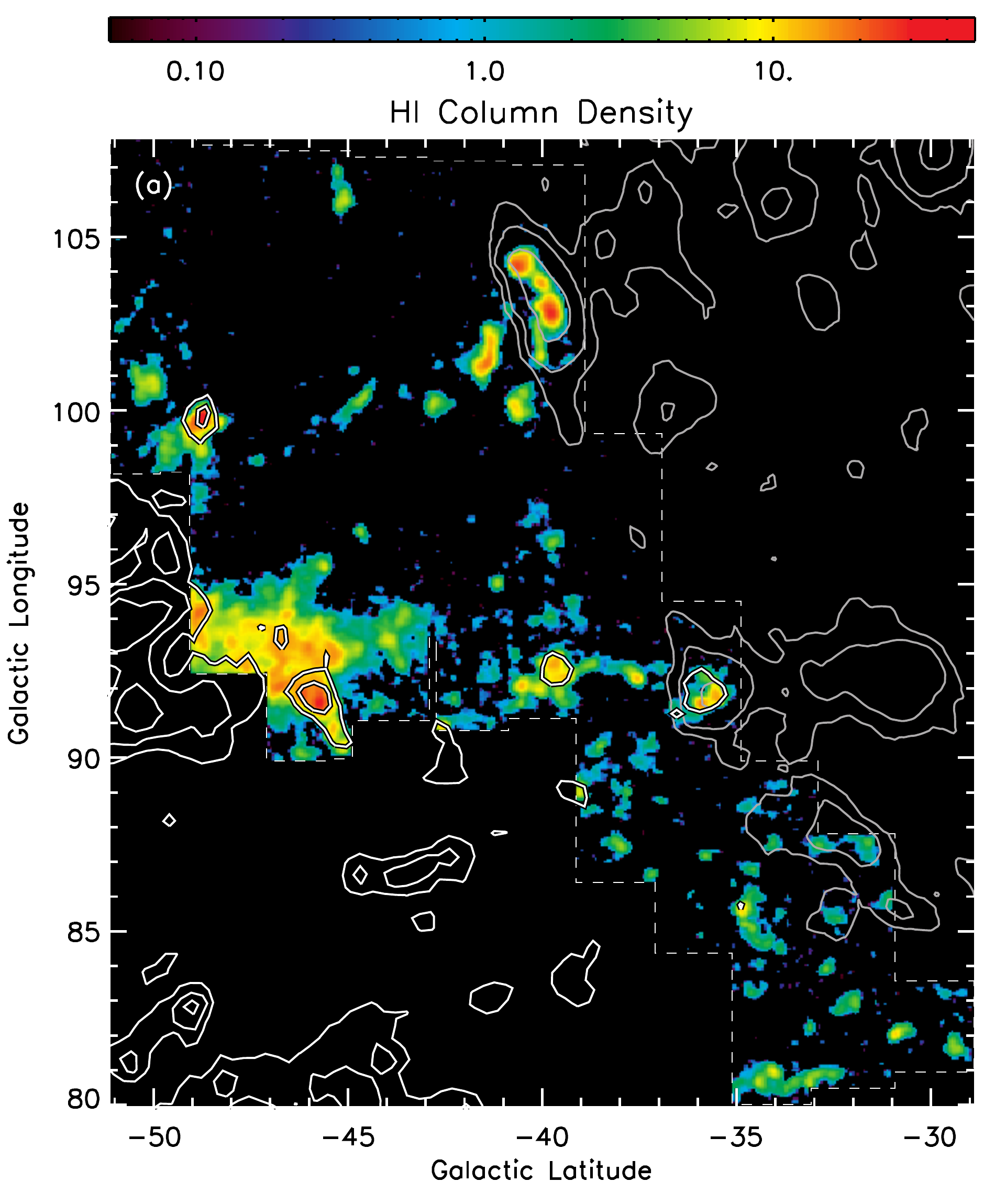} &
\includegraphics[angle=0,scale=0.45]{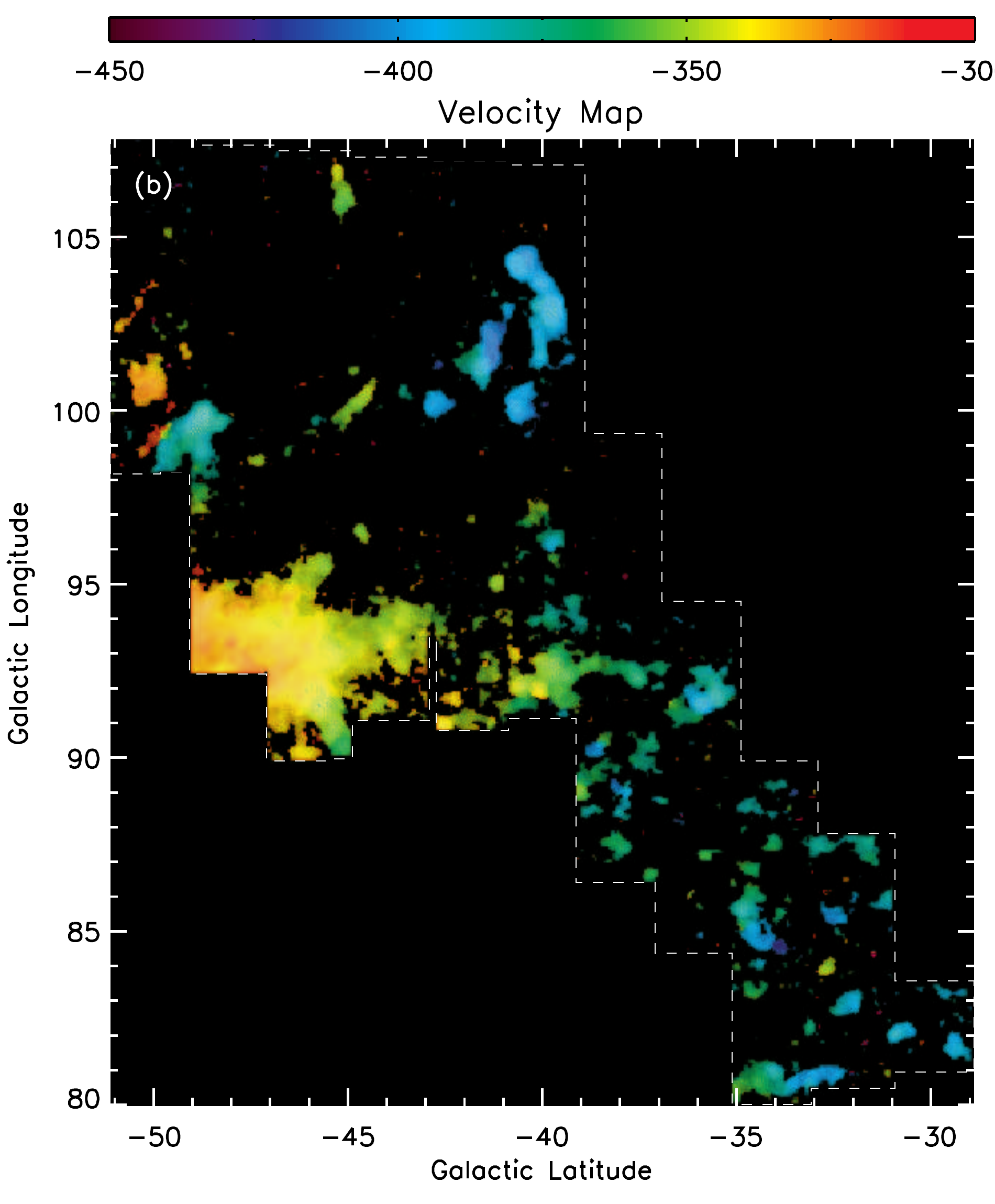}
\end{array}$
\end{center}
\caption{({\em a}) \hi column density, \nhie, (in units of $10^{18}$ atoms cm$^{-2}$) of the MS gas ($-502<$ \vlsr $<-266$ \kmse)
after 3$\sigma$ filtering of the GBT datacube.  The end of the detected MS-tip from the P03 HIPASS data is shown in white contours
(at 0.2, 0.5 and 1$\times10^{19}$ atoms cm$^{-2}$), while gray contours show the MS--like gas in the southern edge of the BT04 survey
(at 0.2, 1, and 5$\times10^{18}$ atoms cm$^{-2}$).
There is nearly-continuous MS emission across our survey at $l\approx93$\dgr
and $l\approx100$\degr.  This shows continuity of the MS in the eastern part of the region (between the ``classical'' MS and the BT04 survey)
where continuity was previously uncertain (see \S \ref{subsec:combined_results}).  The northwestern region of our survey is filled with MS
cloudlets and suggests that the
western part of the MS is also continuous across this region.  ({\em b}) First-moment map for the data in ({\em a}).  Hue indicates the
flux-weighted velocity
(\vlsr in \kms as indicated by the colorbar); intensity indicates \nhie.  The velocity of the $l\approx100$\dgr gas is $\sim$70 \kms lower
than the gas at $l\approx93$\degr.  A dashed line demarcates the boundary of our survey.
The figures have been oriented so that they are aligned with the other MS figures in this paper (e.g., Figs.\ \ref{fig_combined_coldens_velsigcut}
and \ref{fig_combined_lowres_vmap}).}
\label{fig_gbt_coldens_vmap}
\end{figure*}

We used the GBT to conduct an $\sim$200 deg$^2$, 21--cm survey
(proposal ID: GBT06A-066, 102 hours) to study the continuity of the MS in the region between
the ``classical'' MS and the BT04 survey.
The survey area is centered on ($l$,$b$) $\approx$ (94$\degr$,$-40\degr$) and extends ($\sim$28$\degr\times$22$\degr$) in
($l \times b$) (but not completely filled; see Fig.\ \ref{fig_gbt_coldens_vmap}).  Our survey area
was partly chosen to be complementary to the Arecibo survey of Stanimirovi{\'c} et al.
We used the ``On-the-Fly'' mapping mode (scanning in Galactic longitude) to obtain frequency-switched, 21-cm spectral
line data with the Spectrometer back end.  The usable velocity range is $-960<$ \vlsr $<+540$ \kmse, with a velocity
resolution of 0.16 \kmse.  The observations were taken in 49 ``bricks'' of dimension 2.1$\degr\times$2.1\degr.
Each brick consists of a 37$\times$37 array of 4--second integrations, with a 3.5$\arcmin$ spacing between each integration
(which fully samples the GBT 21--cm half-power beam width of 9.2$\arcmin$).  Neighboring bricks were overlapped by
6$\arcmin$ to allow for a consistent calibration across the survey area.
The survey area can be seen in Figure \ref{fig_gbt_coldens_vmap}.

During each observing session calibration data were obtained for the IAU standard positions S7 and S8 \citep{Williams73},
and, for our own secondary standard position, at ($l$,$b$) = (103.0$\degr$,$-40.0\degr$).

\subsection{Data Reduction}
\label{subsec:gbtred}

We used the GETFS program in GBTIDL\footnote{http://gbtidl.nrao.edu/} to calibrate our frequency-switched data
using a basic $T_a = T_{\rm sys}^{\rm ref} \times {\rm (sig-ref)/ref}$ calibration.  Each integration and linear polarization was calibrated
separately.

We used our own special-purpose IDL routines to perform the baseline fitting
and removal.  Before baseline removal, the data were binned 10$\times$ (giving a velocity resolution of $\sim$1.6 \kmse)
to decrease the noise.  The baseline was fitted for each integration and polarization
separately.  For the XX polarization, a 5th-order polynomial was fitted to the 21-cm spectrum but excluding the Galactic
emission at $-140 <$ \vlsr $< +100$ \kmse.  The fitting process was done iteratively so that emission lines
(anything above 2$\sigma$ in a Gaussian smoothed version of the spectrum) could be excluded from the fit.  The procedure
for the YY polarization was similar although it included a sinusoidal component in order to remove a standing wave with
a period of $\sim$1.6 MHz
in the GBT data (this is explained in more detail in \S \ref{subsec:standingwave}).

After baseline removal, the median (calculated from an emission--free region of the spectrum) was subtracted from each
polarization and then the two polarizations were averaged to obtain the final, reduced spectrum for that
position.  The RMS noise is $\sim$0.072 K per 1.6 \kms channel; our 3$\sigma$ sensitivity is
$\sim$4.9$\times10^{18}$ cm$^{-2}$ over 20 \kmse.

All 49 bricks were regridded onto a single datacube with a Galactic cartesian grid and a 3.5$\arcmin$ spacing
using the IDL routine TRIGRID with linear interpolation (each velocity channel independently).
The datacube was then Gaussian smoothed
in velocity (FWHM = 16 \kmse) and spatially (FWHM = 10.5$\arcmin$), giving a final RMS noise level of
$\sim$0.0075 K per 1.6 \kms channel.  Our 3$\sigma$ sensitivity is $\sim$5.2$\times10^{17}$ cm$^{-2}$ over 20 \kms
in the final, smoothed datacube.

We found and removed a persistent but weak ($\sim$5mK) emission line in our datacube.  We call this feature the GBT
\hi ``ghost'' and find that it is spurious and likely due to an improperly filtered sideband on a local oscillator at
the GBT.  A more detailed description is given in \S \ref{subsec:gbtghost}.

\begin{figure*}[ht!]
\begin{center}$
\begin{array}{cc}
\epsscale{1.00}
\includegraphics[angle=0,scale=0.415]{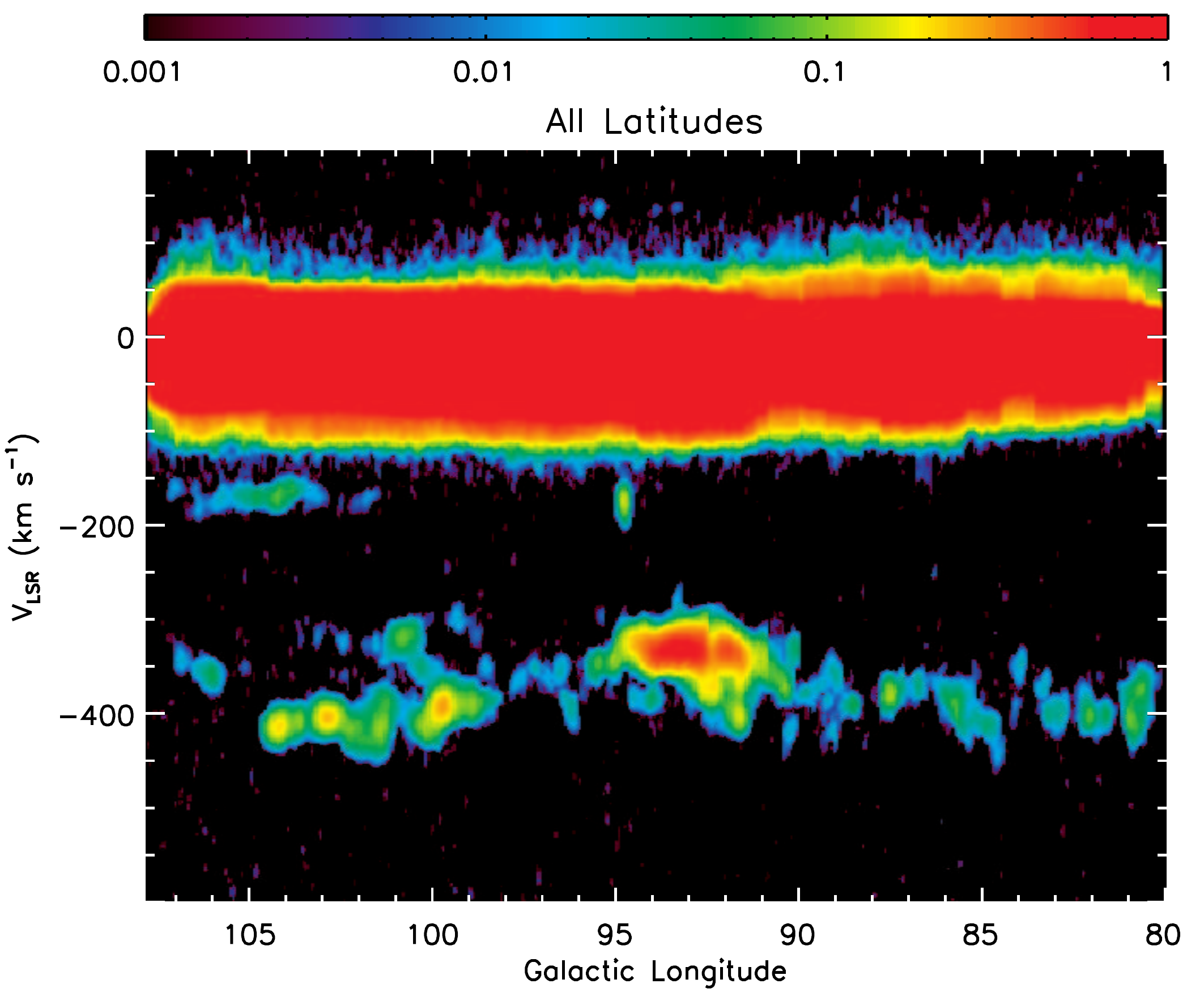} &
\includegraphics[angle=0,scale=0.415]{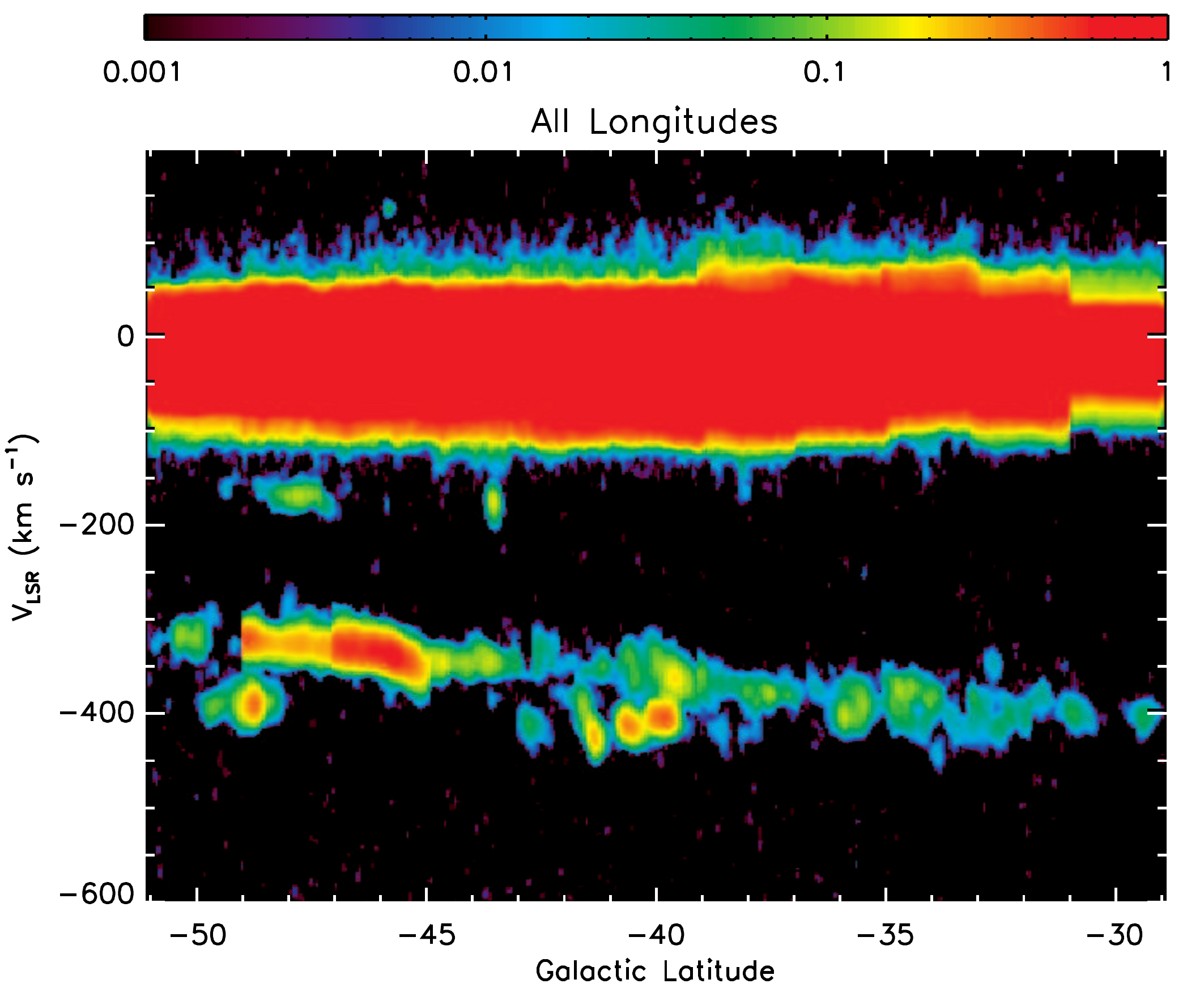}
\end{array}$
\end{center}
\caption{({\em a}) Integrated intensity of the datacube along $b$ (in units of K deg) after $\sigma$-filtering.
The velocity of the $l\approx100$\dgr gas (\vlsr$\approx-406$ \kmse) is $\sim$70 \kms lower than the gas at $l\approx 93$\dgr
(\vlsr$\approx-337$ \kmse).
({\em b}) Integrated intensity of the datacube along $l$ (in units of K deg) after $\sigma$-filtering.  The velocity
gradient of the MS with $b$ is clearly apparent (since $b$ is nearly parallel with the MS at this location).  Some of
the $l\approx100$\dgr cloudlets can be seen slightly below the main linear trend of MS gas at $b\approx-49$\dgr and
$-40$\degr.}
\label{fig_gbt_velglon_velglat}
\end{figure*}

\subsection{GBT Survey Results}
\label{subsec:gbtresults}

Figures \ref{fig_gbt_coldens_vmap} and \ref{fig_gbt_velglon_velglat} show our final, smoothed GBT datacube.
An ``RMS map'' that contains a robust measure of the RMS noise for each position was also created (not shown).
To pull out real features the data were $\sigma$-filtered so that only pixels in the datacube that were
above 3$\times$ the RMS noise (for their respective position) were used to make the figures.

The column density of the MS gas ($-502<$ \vlsr $<-266$ \kmse) is shown in Figure \ref{fig_gbt_coldens_vmap}a.
Contours of the end of the MS from the P03 HIPASS data
(white) and the MS--like gas in the southern edge of the BT04 survey (gray) reveal the region where the continuity
of the MS is unclear and 
that is filled by our GBT survey (mostly on the eastern part)\footnote{In this region of the sky the celestial and
galactic coordinate systems are nearly aligned and, therefore, references such as ``eastern'' or ``northwestern''
can refer to either coordinate system.}.  The HW88 data detect the MS in a wedge shaped region in the
central part of this figure (Fig.\ \ref{fig_prevsurveys}).
There is nearly-continuous MS emission across our GBT survey at $l\approx93$\dgr which confirms the continuity of
the MS in this region, as first seen by HW88, and reveals it to be composed of many small clumps and cloudlets.
Additionally, there is nearly-continuous MS emission across our survey at $l\approx100$\dgr that connects the
``classical'' MS to the BT04 survey for the eastern ($l\approx100$\degr) portion of the MS.  Previous data (HW88, P03)
did not show continuity of the MS in this eastern region.
The northwestern
region of our survey is filled with cloudlets at MS velocities and suggests that the western portion is also continuous
across this region.
Figure \ref{fig_gbt_coldens_vmap}b shows the velocity map for the MS gas.
The gas at $l\approx100$\dgr is at slightly lower velocity (more negative) than the gas at $l\approx93$\degr.
Figure \ref{fig_gbt_velglon_velglat}a shows the integrated intensity of the datacube along $b$ and
Figure \ref{fig_gbt_velglon_velglat}b the integrated intensity along $l$. The velocity
gradient of the MS with $b$ is clearly apparent (since the MS nearly follows a constant line of Galactic longitude at this location).
Some of the $l\approx100$\dgr cloudlets can be seen slightly below the main linear trend of MS gas at $b\approx-49$\dgr and
$-40$\degr.

In summary, our dataset shows that the MS is continuous across this region between the end of the ``classical'' MS and the
BT04 survey.  Therefore, the MS is much longer than previously known with confidence.
In the next section we combine our data with other \hi
datasets in order to obtain a complete picture of the tip of the MS and to determine the full extent and distribution
of the MS on the sky and in velocity.

\section{The Combined Magellanic Stream Tip Datacube}
\label{sec:combined}

To gain a clear picture of the entire tip of the MS we combined our GBT datacube with
(1) the S08 Arecibo datacube, (2) a part of the \citet[][hereafter Br05]{Br05} Parkes datacube,
and (3) the BT04 Westerbork datacube.

\subsection{Combining the GBT, Arecibo, Parkes, and Westerbork Datacubes}
\label{subsec:combined_combining}

Our combined MS--tip datacube is on a Cartesian grid on the MS coordinate
system of \citet{Nidever08}\footnote{Note that this coordinate system was designed to bisect the Magellanic Stream
and differs from the ``Magellanic coordinate
system'' used by \citet{Wakker01} and P03.} with an angular step size of 3\arcmin,
a velocity step of 1.5 \kmse, and a velocity range of $-548.5\leq$ \vlsr $\leq -100.0$ \kmse.
A cubic spline was used to interpolate the datacubes onto the new velocity scale and the IDL function TRIGRID
was used to regrid the datacube spatially one velocity channel at a time.  Each dataset was
regridded onto the new MS--tip grid separately before all datasets were combined.  

\vspace{5pt}
\noindent
{\em GBT Data:}  The raw GBT data have an angular resolution of 9.2\arcmin, angular sampling of 3.5\arcmin, and
velocity spacing of 0.16 \kms (before binning).  The original, velocity-binned, GBT data were gridded onto the new
datacube, and then smoothed with a Gaussian (FWHM$=$15 \kmse) in velocity and spatially (FWHM$=$6\arcmin).
The average final RMS noise is 0.011 K.

\vspace{5pt}
\noindent
{\em Arecibo Data:}  The S08 Arecibo datacube is a combination of three observing programs with
differing $S/N$.  The original data have an angular resolution of 3.5\arcmin, angular sampling of 1.8\arcmin, and
velocity spacing of 0.18 \kmse, but after gridding have an angular spacing of 3\arcmin~and velocity spacing of 1.47 \kms
(after binning).  We masked out regions of the datacube with missing or bad data, as well as regions with poor
$S/N$ that overlapped the GBT data (mainly in the northern part).  After the Arecibo datacube was regridded onto
our MS--tip datacube it was smoothed with a Gaussian (FWHM$=$15 \kmse) in velocity.
The average final RMS noise is 0.012 K.

\vspace{5pt}
\noindent
{\em Parkes Data:} The Br05 Parkes data have an angular resolution of 14.1\arcmin, angular sampling
of $\sim$5\arcmin, velocity spacing of 0.82 \kmse, and cover a velocity range of
$-410 \lesssim$ \vlsr$\lesssim380$ \kmse, but after gridding have an angular step of 10\arcmin~with the same
velocity spacing.  After the Parkes datacube was regridded onto our
MS--tip datacube it was smoothed with a Gaussian (FWHM$=$15 \kmse) in velocity.
The average final RMS noise is 0.032 K.

\vspace{5pt}
\noindent
{\em Westerbork Data:} The BT04 Westerbork auto-correlation data have an angular resolution of $\sim$50\arcmin~
(effective telescope beam), an angular sampling of 15\arcmin, and velocity spacing of 8.3 \kms (Hanning smoothed
to give a spectral resolution of 16.6 \kmse).  The gridded datacube was resampled onto the Local Group Standard of Rest
(LGSR) velocity system with a velocity range of $-535 \leq$ \vlgsr $\leq 95$ \kms ($-700 \lesssim$ \vlsr $\lesssim -50$ \kmse).
After converting the velocities to \vlsre, we discovered that there was a systematic velocity offset 
between the Westerbork data and the GBT and Arecibo data where they overlapped.  We used the
Leiden-Argentine-Bonn (LAB) \hi all-sky survey \citep{LAB}
to determine the velocity offset in all positions that had emission detected in
both the LAB and Westerbork data.  The velocity offset changed systematically with position with an average offset of
$\sim$30 \kmse.  The additive offset (i.e.~$v_{\rm correct}=v_{\rm Westerbork}+v_{\rm offset}$) was well-fitted by

\[ v_{\rm offset} = 58.24 + 0.1954 \alpha - 0.00959 \alpha^2 - 0.7435 \delta \]


\noindent
where $\alpha$ and $\delta$ are in degrees.  We applied this velocity offset to the Westerbork data and converted to \vlsr
before regridding onto our MS--tip datacube.  Since M31 and M33 are very prominent in the Westerbork datacube and would
``contaminate'' our MS position-velocity diagrams, we blanked their emission from the datacube.
The average final RMS noise is 0.0012 K.

\begin{figure*}[ht!]
\begin{center}
\epsscale{1.00}
\includegraphics[angle=0,scale=0.50]{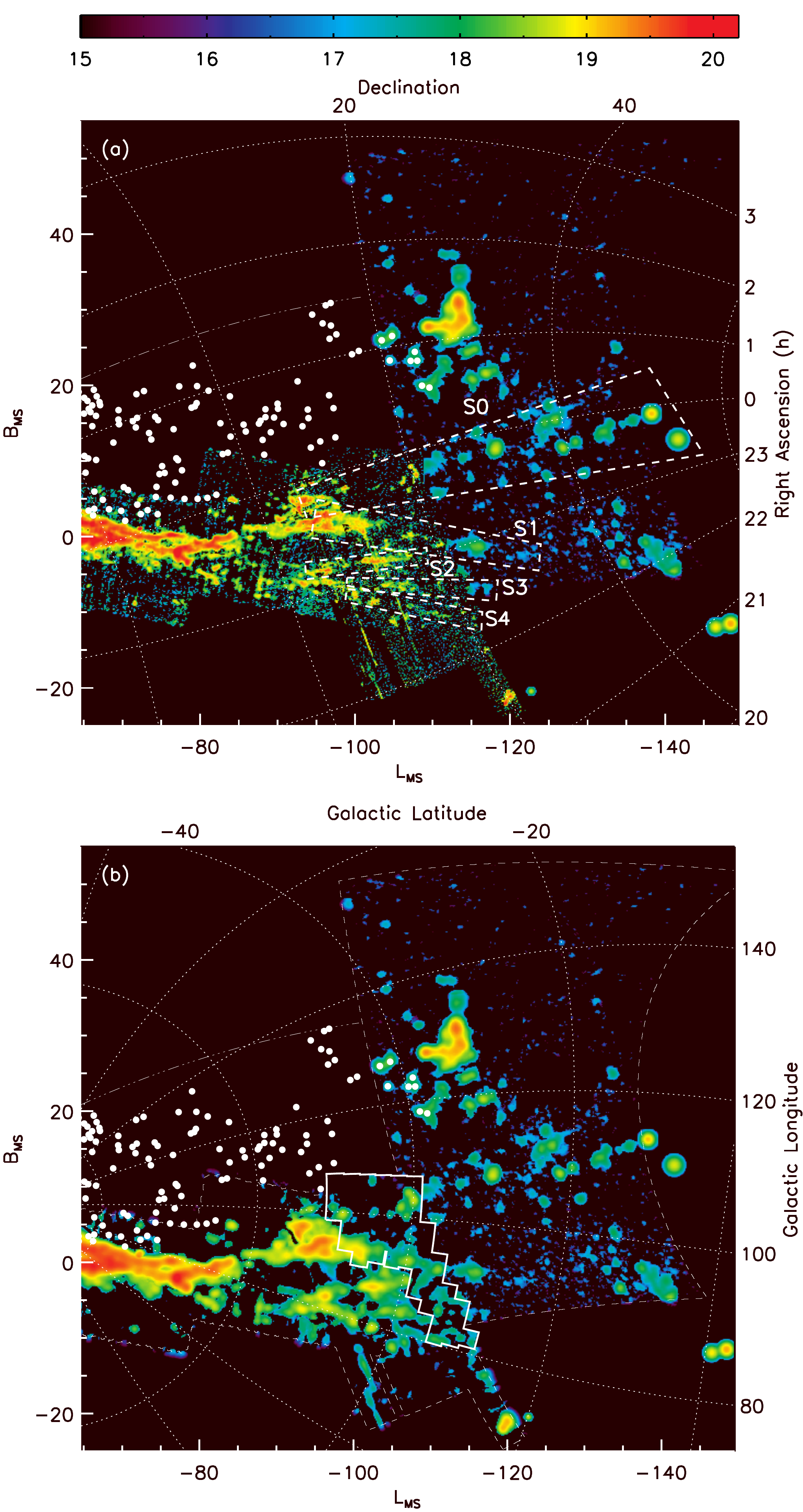}
\end{center}
\caption{\hi column density, \nhie, of the MS--tip combined datacube ($\log$(\nhie) in units of atoms cm$^{-2}$)
after an MS velocity selection and 3$\sigma$ filtering have been applied.  ({\em a}) The ``hi-res'' datacube with all
four datasets (GBT, Arecibo, Parkes, and Westerbork) in their native resolutions.  A celestial coordinate grid
is overlaid. The various MS filaments described in \S \ref{subsec:combined_results} are outlined with dashed lines.  S1--S4 were
defined in S08 and we extend them slightly with the addition of the BT04 data.  We define the new eastern--most
filament as S0; which splits with S1 near \lmse$\approx-95$\dgr but extends until \lmse$\approx-143$\degr.
({\em b}) The ``low-res'' datacube with all datasets at 50\arcmin~resolution.  A Galactic
coordinate grid is overlaid and the boundary of our GBT survey is indicated by a solid white line.
CHVCs from WK08 that they associated with the MS are represented by white
dots.  A dashed line indicates the boundary of the combined \hi datacubes, and a dashed--dotted line delineates the
eastern edge of the region studied by WK08 (0\h 00\m\ $\leq \alpha \leq$ 1\h 30\m\ and $-30$\dgr $ \leq \delta \leq$ +25\degr).
The MS is seen here to be $\sim$40\dgr longer than the ``classical'' MS (which ends near \lms$\approx-105$\degr), and now extends
to the end of the Westerbork survey at $\delta\approx+50$\degr.  The MS is very complex in this region with many forks
and filaments --- see text for details.
Wright's Cloud is the backward ``L''-shaped cloud at (\lmse,\bmse)$\approx$($-114$\degr,$+29$\degr).}
\label{fig_combined_coldens_velsigcut}
\end{figure*}

We used a rank order of decreasing priority as GBT:Arecibo:Parkes:Westerbork to select which data to use for regions where
the datasets overlap.  We created a ``hi-res'' datacube, where all four datasets have their
original resolutions (Fig.\ \ref{fig_combined_coldens_velsigcut}a), and a ``low-res'' datacube, where the combined GBT/Arecibo/Parkes
data were smoothed to a resolution of 50\arcmin~before being combined with the Westerbork data
(Fig.\ \ref{fig_combined_coldens_velsigcut}b).  In the hi-res datacube the intensity
of the emission in the Westerbork part is generally low (by $\sim$2$\times$) due to beam dilution, which creates
an artificial drop in emission/column density at the interface between the Westerbork and GBT/Arecibo data.  The 
low-res datacube gives generally smoother images because the data are relatively homogeneous in angular resolution.

\begin{figure}[t]
\includegraphics[angle=0,scale=0.40]{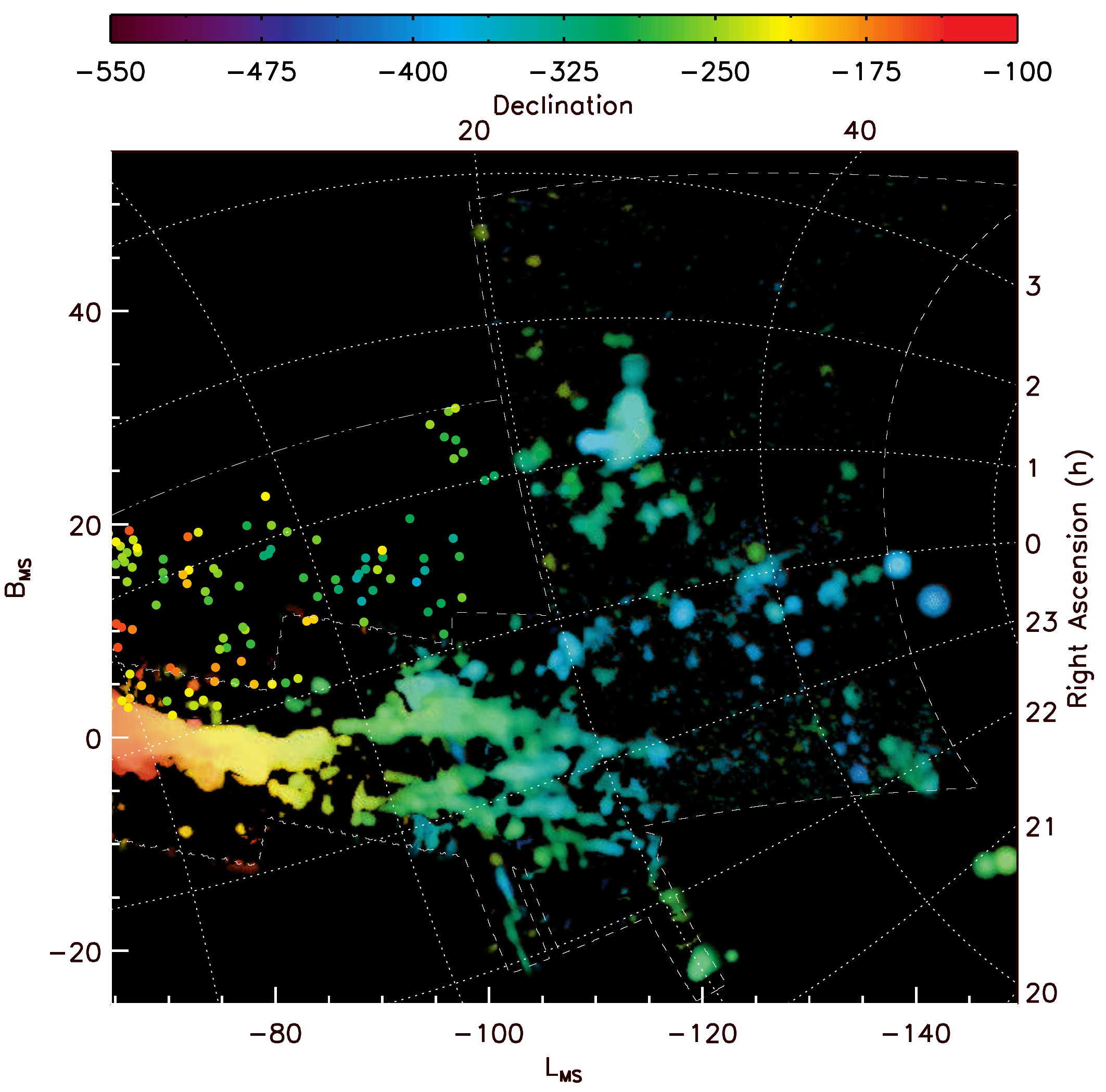}
\caption{Velocity map of the 50\arcmin--smoothed MS tip datacube.  Hue indicates the flux weighted velocity
(\vlsr in \kms as indicated in the colorbar) and intensity indicates \nhie.  The colored dots are CHVCs from WK08.
The velocity gradient along the MS is clear from the color coding.}
\label{fig_combined_lowres_vmap}
\end{figure}

\begin{figure*}[t]
\includegraphics[angle=90,scale=0.64]{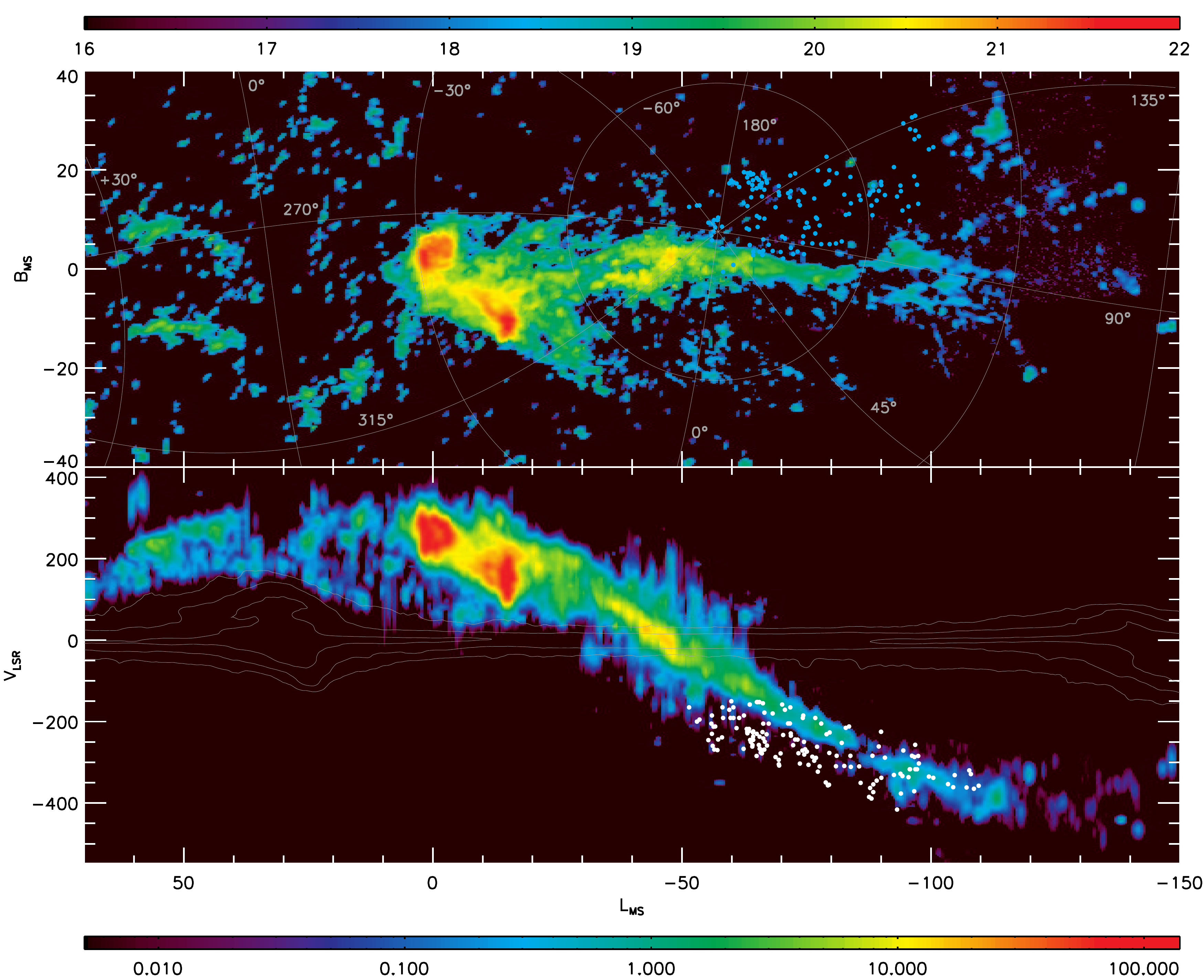}
\caption{Combined data for all the Magellanic \hi gas. ({\em Top}) \hi column density ($\log$(\nhie) in units of
cm$^{-2}$).  The MS--tip datacube is used for \lmse$\lesssim-65$\dgr and the LAB MS Gaussians from
\citet{Nidever08} are used for the rest.  The CHVCs from WK08 are shown as light-blue dots (not color-coded).
({\em Bottom}) Total intensity of the Magellanic \hi integrated along \bms (in units of K deg).  The MS Gaussians are shown
with their true width and not just at their central velocity as in \citet{Nidever08}.  The CHVCs from WK08 are shown
as white dots. The Galactic \hi emission is shown as gray contours (at 10, 100, and 1000 K deg).}
\label{fig_allms_2panel}
\end{figure*}

We have also identified five Compact or extended High Velocity Clouds (C/HVCs)
from the catalog of \citet{deHeij02} (numbers 237, 304, 307, 391, 402 in their Table 1) that lie outside the area covered by the four
datasets in our MS--tip datacube but might be associated with the MS because they match the MS in position and velocity.
The C/HVCs were
added into the combined MS--tip datacubes with the structural \hi parameters given by de Heij et al.  Since the Leiden/Dwingeloo \hi Survey
\citep[LDS;][]{LDS} was used to identify these \hi clouds, they have an angular resolution of 36\arcmin~and velocity
resolution of $\sim$1 \kmse.  Most of the CHVCs of \citet[][hereafter WK08]{Westmeier08}, which they identified as being associated with the
MS, also lie outside the area surveyed by the four datasets.  However, structural parameters are not available for these CHVCs
and therefore we did not add them to our MS--tip datacube but instead plot them as dots in
Figures \ref{fig_combined_coldens_velsigcut}--\ref{fig_allms_2panel}.

\subsection{MS Tip Results}
\label{subsec:combined_results}

\begin{figure}[ht!]
\includegraphics[angle=0,scale=0.69]{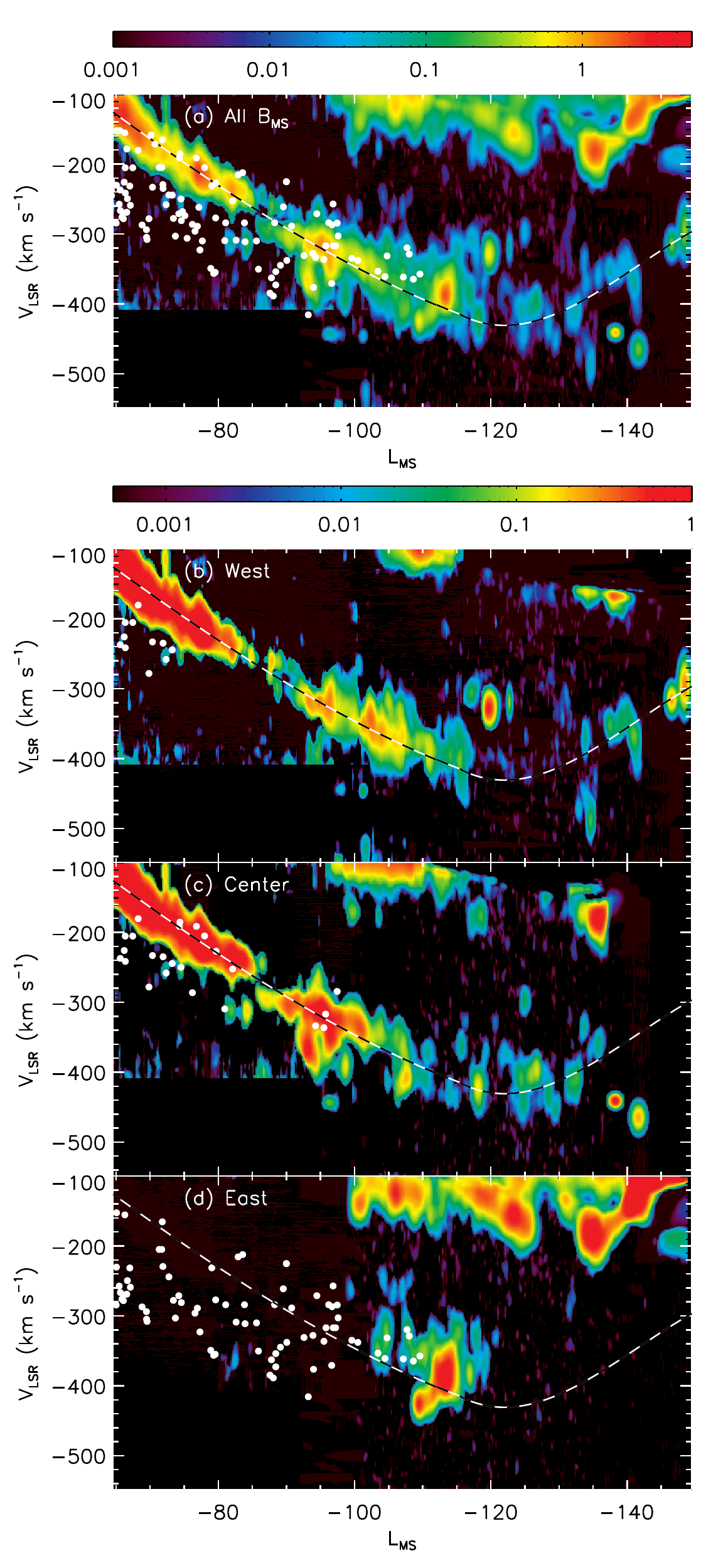}
\caption{({\em a}) Total intensity of the 50\arcmin--smoothed low-res MS--tip datacube integrated along \bms (in units of K deg on
a logarithmic scale).  There is a velocity inflection near \lms$\approx-120$\dgr where the velocity flattens
out and begins to increase.  White dots are the CHVCs from WK08, and a dashed fiducial line (same in all panels)
shows the proposed general velocity trend.  Wright's Cloud is the red blob at (\lmse,\vlsre)$\approx$($-114$\degr,$-380$ \kmse).
({\em b}) Same as ({\em a}) but only for the western portion of the MS--tip datacube (\bms$\lesssim$ 0\degr) showing
the S2--S4 filaments and part of S1.
({\em c}) Same as ({\em a}) but only for the central portion of the MS--tip datacube
(0\dgr $\lesssim$ \bms $\lesssim$ $+15$\degr) showing filament S0 (and parts of S1).
Two of the \citet{deHeij02} CHVCs are at \lms$\approx-140$\degr.
({\em d}) Same as ({\em a}) but only for the eastern portion of the MS--tip datacube (\bms$\gtrsim+15$\degr) showing
the clouds around Wright's Cloud and many of the WK08 CHVCs.  The sudden cutoff in velocity at \vlsre=$-410$
\kms for \lms$\gtrsim$$-92$\dgr in all panels is due to the velocity limit of the Br05 Parkes datacube.}
\label{fig_combined_lowres_velmlon_4panel}
\end{figure}

The \hi column density map of the hi-res/low-res data can be seen in Figure \ref{fig_combined_coldens_velsigcut}
after an MS velocity selection (removing the well-separated local Galactic and intermediate velocity gas -- see
Fig.\ \ref{fig_combined_lowres_velmlon_4panel}a) and
3$\sigma$ filter have been applied.  With the
addition of the GBT and Arecibo data it is now clear that the entire MS
{\em is continuous} with the emission identified by BT04 as an extension of the MS (starting at $\delta$=$+20$\degr).
The MS is at least $\sim$140\dgr long
and thus is $\sim$40\dgr longer than previously established with certainty.  The entire MS system --- including the LA --- is therefore
at least $\sim$200\dgr long.  The MS extends to the limit of the MS--tip datacube (the Westerbork data) at $\delta\approx+50$\dgr and
therefore is plausibly even longer than shown here.  In fact, there is even one HVC from \citet{deHeij02} (number 424)
that is consistent in position and velocity with being an extension of a filament on the eastern side of the MS at \lmse$\approx-163$\dgr
(see Fig.\ \ref{fig_combined_lowres_extmstip} below).

The \hi data of the MS--tip (Fig.\ \ref{fig_combined_coldens_velsigcut}) show a complex richness of patterns that look like
forks and multiple filaments.  The MS splits and diverges into two filaments near \lms$\approx-80$\degr, which were already visible
in previous data (HW88, P03), but there are more forks farther along the Stream.
S08 noted that the western region of the MS--tip separates into four thin
filaments that they named S1--S4.
The four S08 filaments are indicated in Figure \ref{fig_combined_coldens_velsigcut}a and are extended slightly now that the
BT04 survey data are included.  S3 and S4 appear to run off the edge of the survey coverage, while S1 and S2 are likely converging or
overlapping for \lms$\lesssim-108$\degr.

A new filament of the MS (on the eastern side) is visible in our data that separates from S1 near
\lms$\approx-95$\degr.  Following the S08 nomenclature we call this filament S0.  It is made up of many cloudlets, especially right
after it separates from S1, but the continuity of S0 is clear in our GBT data (Fig.\ \ref{fig_gbt_coldens_vmap}).  As can be
seen in Figure \ref{fig_combined_coldens_velsigcut}, S0 extends all of the way to the end of the coverage area near \lmse$\approx-140$\degr.
It deviates from the equator of the MS coordinate system and extends diagonally at an angle of $\sim$14.6\dgr (slope of $\sim$0.26).
The S0 filament also follows a great circle with a north pole of ($l$,$b$)=(205.2\degr,$-15.4$\degr)
(Fig.\ \ref{fig_combined_lowres_extmstip}; black--white dashed line)
which is $\sim$18\dgr from the north pole of the MS coordinate system ($l$,$b$)=(188.5\degr,$-7.5$\degr).
The western portion of the MS, which is probably a combination of S1, S2, and possibly S3\footnote{S4 runs off the coverage near
(\lmse,\bmse)=($-116$\degr,$-11$\degr).}, also extends to the end of the coverage area near \lmse$\approx-140$\degr.
More data are needed in the west and north to track the MS filaments further.

Figure \ref{fig_combined_lowres_vmap} shows the velocity map of the low-res datacube with the WK08
CHVCs plotted as colored dots.  The strong velocity gradient ($\sim$6 \kms deg$^{-1}$) with \lms is apparent from the color coding.
Figure \ref{fig_combined_lowres_velmlon_4panel} shows various \vlsre--\lms
position-velocity diagrams of the MS--tip low-res datacube.  Figure \ref{fig_combined_lowres_velmlon_4panel}a shows the
total intensity of the entire datacube integrated in \bmse.  The strong velocity gradient with \lms is again apparent, and
so is a velocity inflection near \lmse$\approx-120$\dgr where the velocity of the MS levels off and then
starts to increase.  This is the first evidence of a leveling-off of the MS velocity anywhere along its length.

Figure \ref{fig_combined_lowres_velmlon_4panel}b is a \vlsre--\lms diagram similar to Figure
\ref{fig_combined_lowres_velmlon_4panel}a, but only for the western portion of the datacube (\bms $\lesssim$ $0$\degr)
showing the velocity behavior of the S2--S4 filaments and part of S1.  Most of the gas follows the velocity
inflection (and the velocity fiducial line) including the low column density gas at $-132$\degr$\lesssim$ \lms $\lesssim-117$\degr.
There is a cloud at (\lmse,\vlsre)$\approx$($-135$\degr,$-480$ \kmse) that is consistent with a near-linear
extrapolation of the MS, and that might indicate that not the entire MS participates in the velocity inflection.
The clouds at (\lmse,\vlsre)$\approx$($-120$\degr,$-320$ \kmse), above the main MS in the figure, are in the far western part of the
datacube at \bmse$\approx-20$\dgr and are not part of the main MS filaments (S0--S4).

A \vlsre--\lms diagram of the central portion of the datacube (0\dgr $\lesssim$ \bms $\lesssim$ $+15$\degr) including
filament S0 (and parts of S1) is shown in Figure \ref{fig_combined_lowres_velmlon_4panel}c.  The velocity levels out
and then increases, but not as quickly as for the western portion of the MS (Fig.\ \ref{fig_combined_lowres_velmlon_4panel}b).
The velocity scatter for the central region is
larger than for the western.  There are two clouds near (\lmse,\vgsre)$\approx$($-140$\degr,$-450$ \kmse) that might not
participate in the velocity inflection but continue to more negative velocities.
Figure \ref{fig_combined_lowres_extmstip} extends the \lms range of
Figure \ref{fig_combined_lowres_velmlon_4panel}c to include HVC 424 from \citet{deHeij02} which appears to be an
extension of the S0 filament at \lmse$\approx-163$\degr.
This cloud, as well as the clouds at \lmse$\approx-140$\degr, are consistent with a possible shallower velocity
inflection for the S0 filament than for the rest of the MS.

\begin{figure}[t]
\includegraphics[angle=0,scale=0.50]{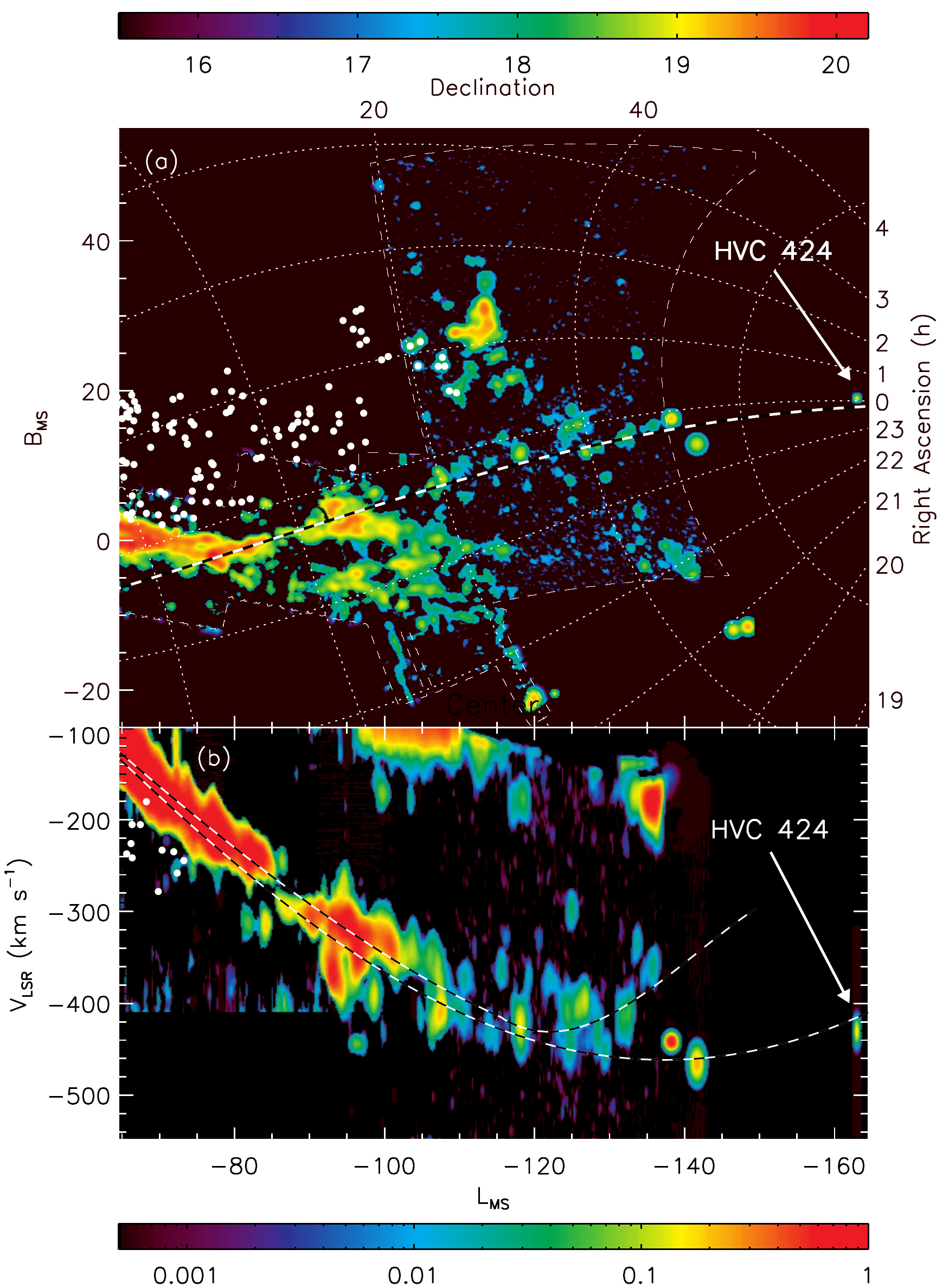}
\caption{({\em a}) Same as Fig.\ \ref{fig_combined_coldens_velsigcut}b but with an extended
\lms range and showing HVC 424 from \citet{deHeij02} at \lmse$\approx$$-163$\dgr that might belong to 
an extension of the S0 filament of the MS.  The black--white dashed line indicates the great circle with north
pole of ($l$,$b$)=(205.20\degr,$-15.40$\degr) that best fits the S0 filament of the MS and which also passes
close to HVC 424.  The great circle also fits the eastern portion of the MS (S0+S1 before they split) all the way
to \lms$\approx-80$\degr.
({\em b}) A similarly extended version of Fig.\ \ref{fig_combined_lowres_velmlon_4panel}c.
A second \vlsre--\lms fiducial is shown that fits HVC 424 .}
\label{fig_combined_lowres_extmstip}
\end{figure}

Finally, Figure \ref{fig_combined_lowres_velmlon_4panel}d shows the part of the datacube to the east of the S0 filament
(\bms$\gtrsim+15$\degr) which includes Wright's Cloud \citep{Wright79} at \lms$\approx$ $-114$\dgr and most of the
WK08 CHVCs.  The clouds around
Wright's Cloud appear to be an extension of the WK08 CHVCs and they all generally follow the MS velocity
trend.  Some authors (Wright 1979, BT04, Putman et al.\ 2009) have suggested that Wright's Cloud
might be associated with the MS; the velocity
agreement of the MS and Wright's Cloud in Figures \ref{fig_combined_lowres_velmlon_4panel}a and d makes this association
likely.  We discuss this possibility further in \S \ref{sec:discussion}.

\begin{figure*}[ht!]
\begin{center}
\includegraphics[angle=0,scale=0.62]{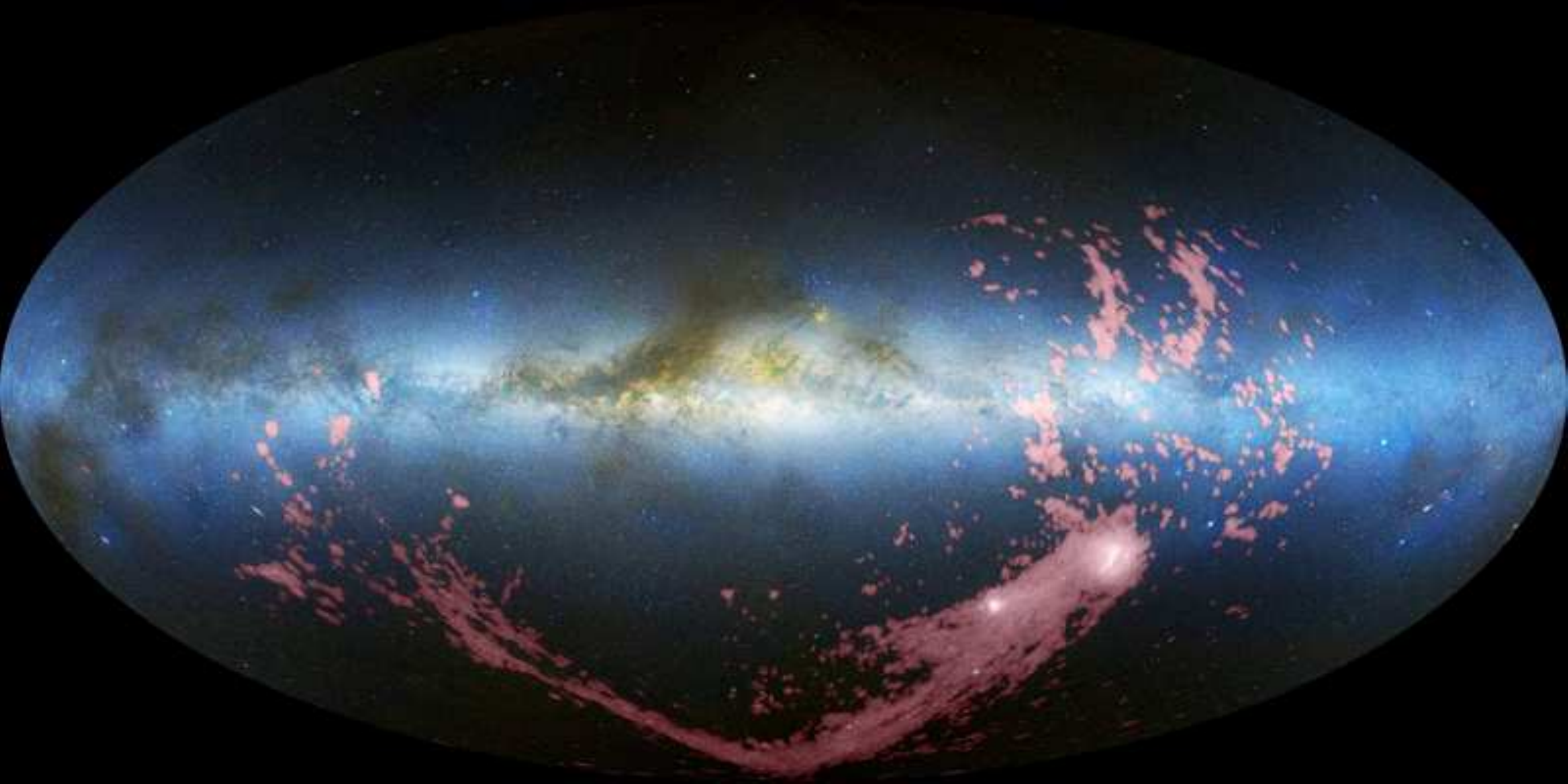}
\end{center}
\caption{The Magellanic Stream and Clouds in \hi (red) with an optical all-sky image
\citep[blue, white, and brown;][]{Mellinger09} in Aitoff projection with the direction to the Galactic center at the center.}
\label{fig_msmw}
\end{figure*}

To calculate the mass of the newly-confirmed part of the MS-tip we remove
all of the \hi emission that was previously observed (i.e., the ``classical'' MS) by subtracting the Br05 Parkes data from
our combined MS--tip column density map.  Orbits of the MCs and simulations of the MS indicate that
the distance increases very rapidly at the tip of the MS especially for \lms$\lesssim-$100\dgr \citep[e.g.,][]{GN96,C06,Besla07,Mastro09},
and the newer orbits and simulations based on the new $HST$ proper motions of the MCs \citep{Kalli06a,Kalli06b} give
significantly larger distances than the older calculations.  Based on the modeling literature we pick a value of 120 kpc as
an approximate distance for the MS-tip ($-$140\degr$\lesssim$\lmse$\lesssim$$-$100\degr).
Using this constant distance for all the gas in the newly-confirmed $\sim$40\dgr portion
of the MS-tip gives a mass of $1.8\times10^7$ \msun $(d/120~{\rm kpc})^2$ without Wright's Cloud and $2.8\times10^7$ \msun $(d/120~{\rm kpc})^2$
with Wright's Cloud.  Since the mass of the ``classical'' MS is $\sim$$5\times10^8$ \msun (Br05), the mass of the new portion
of the MS--tip corresponds to a $\sim$4\% increase in the total mass of the MS.

As previously mentioned, simulations of the
MS indicate that the distance increases rapidly at the MS-tip and, therefore, using a constant distance for the new extension
is not very realistic.  If the distances (as functions of \lmse) from the K1 and GN96 orbits described by \citet{Besla07} are used,
then the derived mass, without Wright's Cloud, increases to $5.8\times10^8$ \msun and $3.0\times10^7$ \msun respectively,
and $6.3\times10^8$ \msun and $4.1\times10^7$ \msun with Wright's Cloud.
The K1 masses are by far the largest because the high-velocity orbit reaches large distances ($>1$ Mpc) for \lms$<-125$\degr.
Due to ram pressure, the unbound MS gas will slow down, lose angular momentum, and fall in its orbit around the MW.\
Therefore, the actual MS distances are likely to be smaller than those from the K1 orbit (which is based on the most
accurate proper motion measurements of the MCs), and a more realistic (and conservative) mass estimate is likely to be closer to
$\sim5\times10^7$ \msune.  However, the distances used are poorly constrained by observations and the derived masses are
therefore quite uncertain.

\section{The Entire Magellanic Stream}
\label{sec:allms}

We combined the MS--tip datacube with the LAB MS Gaussians from \citet{Nidever08} to create images of all the detected Magellanic
gas.  Data were included from P03 for small \hi MS cloudets that were not well detected in the LAB survey
($-95$\degr~$\lesssim$~\lmse~$\lesssim-50$\degr, $-30$\degr~$\lesssim$~\bmse~$\lesssim-5$\degr).
The column density, \nhie, of the entire $\sim$200\dgr MS+LA system is shown in Figure \ref{fig_allms_2panel}a.  The
separation of the two main MS filaments at \lmse$\lesssim-80$\dgr is clearly visible. The very low
column density of the MS--tip (\lmse$\lesssim-100$\degr) made it difficult to detect in previous surveys,
but it is clearly evident in Figure \ref{fig_allms_2panel}a.

Figure \ref{fig_allms_2panel}b shows the \vlsre--\lms position-velocity diagram (integrated in \bms
for $-40$\dgr $\leq$ \bms $\leq$ $+40$\degr ; in units of K deg)
for the entire MS.  For most of the length of the MS ($-110$\degr$\lesssim$\lmse$\lesssim$$-$30\degr) its velocity follows
a fairly linear gradient ($\sim$7.3 \kms deg$^{-1}$); this changes, however, at the very tip where the velocity levels out
and starts to increase (\lmse~$\lesssim$$-$110\degr).

We can now see the WK08 CHVCs (light blue dots in Fig.\ \ref{fig_allms_2panel}a) and the cloudlets
at the far eastern part of the MS--tip datacube (near Wright's Cloud) in the context of the entire MS.  The WK08
CHVCs appear to be an extension of the cloudlets near (\lmse,\bmse)$\approx$($-45$\degr,$+15$\degr), just above the main body of
the MS, which themselves are probably an extension of the \hi filament emanating diagonally from the center of the Magellanic
Intercloud Region (ICR) at (\lmse,\bmse)$\approx$($-9$\degr,$-2$\degr).  It appears as though all of these cloudlets are
connected into one large filamentary distribution that is parallel to, but above in \bmse, the main body of the MS.  This``parallel filament''
has lower column density and is much clumpier than the main MS, and does not appear to continue past Wright's Cloud near \lmse$\approx-120$\degr.
WK08 pointed out that this parallel filament concides with a secondary stream predicted by the numerical simulations
by \citet{GN96}, although it is less apparent or non-existent in more recent models of the MS \citep[e.g.,][]{YN03,C06,Mastro05}.
The existence of MS debris on the other side of the MS (\bms$<0$\degr), which is not seen in the \citet{GN96} simulations, suggests
that the MS is generally much broader once low column densities are included.  Plausibly the process that
is creating the main body of the MS is also liberating small \hi cloudlets that are being spread over a large area of the sky.
It is therefore possible that even more CHVCs in this part of the sky (beyond those discovered by WK08) are part of
this Magellanic Stream ``wake''.

Figure \ref{fig_msmw} shows the entire MS with an optical all-sky image of the Milky Way and the Magellanic Clouds.  The large
extent of the MS on the sky and its orientation with respect to the MW
is revealed in this image, as well as the fact that the observed MS-tip is close to crossing the MW
disk plane.

\begin{figure}[t]
\includegraphics[angle=0,scale=0.38]{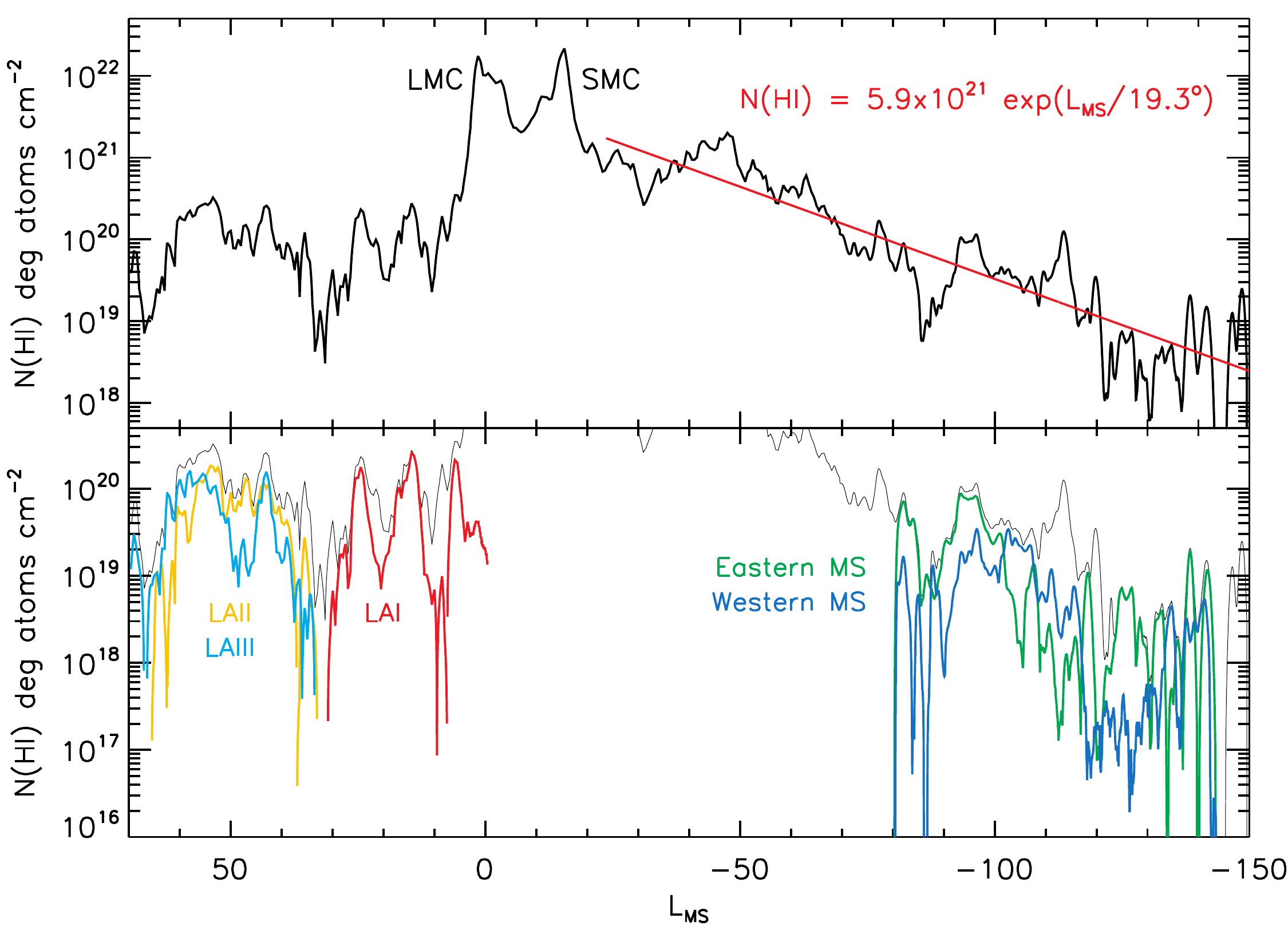}
\caption{Total ``cross--sectional'' column density, \nhie, (integrated in \bmse) of all the Magellanic \hi gas
(same as Fig.\ \ref{fig_allms_2panel}) as a function of \lms (in units of deg atoms cm$^{-2}$).  There is a strong
column density gradient along the MS which is well fit by \nhi$=5.9\times10^{21} \exp(\rm{L_{MS}}$/19.3\degr) deg atoms cm$^{-2}$.
The LA, on the other hand, has a fairly constant column density along its $\sim$60\dgr span.  The lower panel
shows various individual components of the MS and LA: LAI--III, and the eastern (S0+S1) and western (S2--S4) portions of the MS.}
\label{fig_allms_totcol}
\end{figure}

Figure \ref{fig_allms_totcol} shows the column density integrated along \bms for the entire MS+LA system.  There is a strong
column density gradient along the MS as has been previously noted \citep[][P03]{Math74}; this gradient is
well fitted by \nhi$=5.9\times10^{21} \exp(L_{\rm{MS}}$/19.3\degr) cm$^{-2}$.  The LA, on the other hand, has a fairly constant
column density along its $\sim$60\dgr span.  The lower panel shows
various individual components of the MS and LA: LA I--III, and the eastern (S0+S1) and western (S2--S4) portions of the MS.

\section{Discussion}
\label{sec:discussion}

\citet{Besla07} presented new orbits for the MCs that place them much farther from the MW in the past than
previously thought.
If the model orbits are correct, then it is difficult to understand how ram pressure or tidal forces
created a long MS
since these forces require the MCs to be fairly close to the MW for quite some time to be effective.
Some other mechanisms, for example, dynamical processes associated with
star formation, are probably required to help the gas escape the MCs when at large distances from the MW.
The suspected role of star formation in the formation of the MS by Besla et al.\
is in agreement with the independent observational findings of \citet{Nidever08} that one of the MS
filaments originates in a region of the LMC with gaseous outflows plausibly linked to supergiant shells and star formation.
Also, \citet{BC09} found that they
could not reproduce the observed MS distribution in their N-body models if the new space velocities for the MCs were used.
On the other hand, \citet{Mastro09} \citep[and][]{Mastro10} was able to produce a 120\degr--long stream using ram
pressure and the new, higher-velocity
MC orbits.  However, as shown here, the MS is 40\dgr longer than previously established --- and probably
even longer.
It is unclear from current simulations if ram pressure and/or tidal forces alone can account for a stream of this length.
Thus, our finding of a longer MS offers an additional challenge to tidal and ram pressure models.
Future surveys of the MS--tip may provide even more stringent
requirements on MS models and possibly further constrain the MC orbits.

The deviation of the eastern portion of the MS from the equator of the MS coordinate system was already seen by
P03, but the very coherent deviation of the new S0 filament for more than $\sim$45\dgr has not been seen
until now.
The tidal MS models by \citet{C06} reproduce this deviation fairly well (as well as multiple
filaments), while \citet{Mastro05} show a deviation in the opposite direction and \citet{Mastro09}
show only a small deviation (with one filament).  It is not entirely clear why the deviation of the
S0 filament occurs.

\citet{Besla07} show that there is a $\sim$10\dgr deviation of the LMC/SMC orbital paths (to negative \bmse) compared 
to the location of the MS.  Previous MS models \citep[e.g.,][]{GN96,C06,Mastro05} used a small value for the
northern component of the proper motion ($\mu_{N}$) or the (closely-corresponding) X-component of the Galactocentric velocity
($v_x$) in order to bring the MC orbital paths into alignment with the MS.  However, these assumed theoretical values are
inconsistent with the observed proper motions \citep{vdM02,Kalli06a,Kalli06b}.  This disagreement between the orbital paths
and the observed MS cannot be solved by changing the MW mass or by using an aspherical MW potential \citep{Besla07}.  
The newly confirmed
$\sim$40\dgr extension of the MS, and the large deviation of the S0 filament (towards positive \bmse), further exacerbates the
problem.

We investigated the K1 and GN96 LMC orbits from \citet{Besla07} and found that the velocity
inflection does not correspond to any physically identifiable part of these orbit, (unlike, e.g.,
\vgsre=0 \kms which corresponds to peri-- or apo--galacticon).  The inflection is just the point in the orbit at which
the satellite is approaching the observer with the maximum speed, and is then dependent on the
position of the observer with respect to the orbit.  However, at a given observing position, the
position of the velocity inflection (and the maximum approach velocity) is sensitive to the
initial space velocity of the computed orbit.
This newly found, distinct feature in the otherwise remarkably linear velocity structure of the MS
should provide a new constraint for MS simulations.

We have shown that the velocity of Wright's Cloud is in close agreement with the velocity of the MS gas in the
same region of the sky.  This might be an indication that Wright's Cloud is part of the MS.
Another interpretation is that Wright's Cloud is not gas that escaped from the MCs, but rather is
part of a ``Magellanic Group'' of galaxies proposed by \citet{DOnghia08} to have fallen into the MW together.
Since the group is proposed to be quite extended, some objects will fall in before others,
but the whole of the system will enter the MW with similar space velocities.
Wright's Cloud, therefore, could be an \hi cloud (primordial or previously stripped from another galaxy)
that is falling into the MW behind the MCs.  Wright's Cloud should then have a similar space
velocity to the MCs (since they are part of the same group) and will experience similar tidal
and ram pressure forces to the MS that should give rise to a radial velocity very similar to that of the
MS.

As can be seen in Figure \ref{fig_allms_2panel}, MS debris at low column densities is observed at fairly large
angular distances ($\sim$20--30\degr) from the main body of the MS and effectively makes the MS ``wider''.
This implies that the region of the outer gaseous MW halo that has been influenced by the MS is quite large.
The assocation of many CHVCs with the MS (WK08 and this paper) makes it likely that
yet other CHVCs have MS origins as well.
It is conceivable that the new Parkes Galactic All-Sky Survey \citep[GASS;][]{McC09,GASS2} and future surveys,
such as the Galactic Australian SKA Pathfinder survey (GASKAP; J. M. Dickey et al.~2010, in preparation), will
find that the MS is even broader than seen here.

\citet{Nidever08} discovered that at the head of the Stream the two filaments of the MS have sinusoidal
velocity patterns.  One possible explanation is that this is an imprint of the LMC rotation curve on escaping gas.
With this hypothesis Nidever et al.~were able to measure the drift rate of the MS gas away from
the LMC to be $\sim$49 \kms and the age of the $\sim$100\degr--long MS (at 50 kpc) to be $\sim$1.7 Gyr.
Now that the MS is $\sim$40\dgr longer the estimate of the age of the MS must also increase.  Using the
same drift rate the new age estimate for the MS is at least $\sim$2.5 Gyr.
The heliocentric distances of Magellanic orbits
increase significantly at the tail of the MS, which in turn increases our MS age estimate via two effects.
A larger MS distance increases the
linear length of the MS and probably decreases the drift rate (since the ram pressure and tidal forces are weaker at larger
distances), both of which increase the age.  Therefore, the MS is possibly even older than $\sim$2.5 Gyr.

The new MS age estimate of $\sim$2.5 Gyr closely coincides with ``bursts'' of star formation in the
integrated star formation histories of the LMC \citep{Harris04} and SMC \citep{Smecker-Hane02} 
at $\sim$2--2.5 Gyr ago; these nearly simultaneous bursts in each system are suggestive of an interaction between
the MCs at that time. The LMC globular cluster age--gap between 3 and 13 Gyr
\citep[e.g.,][]{DaCosta91,Geisler97,Rich01,Piatti02} indicates a dramatic onset of star formation in the LMC
around $\sim$3 Gyr ago, although such an age--gap is not seen in the SMC clusters \citep{DaCosta91}.
Finally, the orbits by \citet{Besla07} indicate possible encounters of the MCs with each other around 0.2,
3 and 6 Gyr ago.  Therefore, a plausible scenario is that the MCs had a close encounter at $\sim$2.5--3 Gyr
ago that triggered a burst of star formation in the LMC and started the self-propagating supergiant shell blowout of the
MS and LA.

According to the \citet{Besla07} model, the MCs were at a distance of $\sim$600 kpc and well outside the MW
virial radius ($\sim$260 kpc) 2.5 Gyr ago.  Therefore, an age of $\sim$2.5 Gyr for the MS implies that the
MS started forming before the MCs entered the MW.  Ram pressure and tidal forces would be ineffective
at liberating gas at these large distances, and this fact further supports the notion that star formation feedback and/or a close
encounter of the MCs with each other are more likely causes for removing the MS gas at this stage in its evolution.

One point that the ram pressure simulations, tidal simulations, and new LMC orbits \citep{Besla07} agree on
is that the distance from the MW at the MS--tip (beyond $\sim$100\degr) increases quite rapidly (almost exponentially). 
Therefore, the new length of the MS implies that the gas at the MS--tip is much farther away than the
``classical'' MS.  The MS gas we are observing at \lms$\approx130$\dgr is likely at a distance on the order
of $\sim$200 kpc and possibly even larger.  This means that the MS spans a larger range of MW distances than
previously thought, which makes the MS a potentially even more useful probe for constraining the total MW mass and the
MW potential at large distances.

\section{Summary}
\label{sec:summary}

We conducted an $\sim$200 deg$^2$ 21--cm survey with the Green Bank Telescope 
in the region where the continuity of the end of the ``classical'' MS with the MS-like emission reported by
\citet{BT04} was uncertain.
Our survey, in combination with the Arecibo survey by \citet{Stani08}, shows
that the MS gas is in fact both spatially and kinematically continuous across this region.
We made a combined MS--tip datacube incorporating
our GBT datacube, the S08 Arecibo survey, the BT04 Westerbork data, and the \citet{Br05} Parkes
data.  This MS--tip datacube allows us to draw the following conclusions:

\begin{enumerate}
\item The MS is $\sim$140\dgr long, approximately 40\dgr longer than previously verified.  The MS
extends to the edge of the datacube, implying that the Stream may be even longer.  Moreover, there is one
CHVC from \citet{deHeij02} at \lmse$\approx-163$\dgr that is consistent with an extrapolation of
the eastern-most filament of the MS in position and velocity.
\item The tip of the MS is composed of a multitude of forks and filaments.  S08 previously identified filaments
S1--S4 in the western region, and we identify a new filament, S0, on the eastern side.  This filament
splits from S1 near (\lmse,\bmse)$\approx$($-95$\degr,$+3$\degr) and deviates from the equator of the MS coordinate
system for more than $\sim$45\dgr until it reaches the end of the datacube.
\item There is an MS velocity inflection near \lmse$\approx-120$\dgr where the
MS velocity levels out and starts to increase.
\item The mass of the newly extended $\sim$40\dgr at the MS--tip is $\sim$$2\times10^7$ \msun  $(d/120~{\rm kpc})^2$
(including Wright's Cloud increases this by $\sim$50\%) and increases the total mass of the MS by $\sim$4\%.
\end{enumerate}

Five C/HVCs from the catalog by \citet{deHeij02} (numbers 237, 304, 307, 391, 402 in their Table 1) match the
MS in position and velocity and, therefore, might be associated with the MS.  Wright's Cloud also matches the velocity
of the MS and is nearby in position, which makes an association with the MS possible.  The CHVCs that WK080
associated with the MS follow the MS velocity trend in our \vlsre--\lms diagrams and might belong to a
 ``parallel filament'' next to the MS that stretches from the head of the MS to cloudlets near Wright's Cloud.

The total column density (integrated in \bmse) along the MS drops markedly and follows an exponential decline
with the angular distance from the LMC well-fitted by \nhi$=5.9\times10^{21} \exp(L_{\rm{MS}}$/19.3\degr) cm$^{-2}$.

An increased length of the MS also increases the age estimate of the MS.  Using the velocity sinusoidal pattern
of the LMC filament of the MS as a chronometer \citep[as was previously done by][]{Nidever08}, we estimate that the age
of the $\sim$140\degr--long MS is $\sim$2.5 Gyr.  This timescale coincides with bursts of star formation in both
the LMC and SMC as well as a possible close encounter between the MCs that could have triggered the new era of star
formation and the formation of the MS via supergiant shell blowout.

These new observational characteristics of the MS should offer further constraints on
MS simulations and provide data that can be used in future work
to elucidate more accurate dynamical and structural
information for the Magellanic System and its interaction with the MW.

We encourage further \hi observations of the MS--tip to track the MS to ever greater lengths until
its actual end is found.

\acknowledgements

We thank Gurtina Besla, Sne\v{z}ana Stanimirovi{\'c}, Nitya Kallivayalil, Chiara Mastropietro, J\"urgen Ott,
Robert Braun, Tobias Westmeier, and Kenji Bekki for useful discussions and comments.  We also thank
Robert Braun, Snezana Stanimirovi{\'c}, Lister Staveley-Smith, Tobias Westmeier,
Mary Putman, Axel Mellinger, and Gurtina Besla for sending us their data/models.  We are grateful to Jay
Lockman and Toney Minter for their help with the GBT observing and reduction.
We thank the referee for useful comments and suggestions that improved the manuscript.
D.L.N. greatly thanks Amy Reines for her enduring support, suggestions, and guidance.
D.L.N. is supported by the Green Bank Telescope Student Support Program, the ARCS Foundation,
a University of Virginia President's Fellowship, the Virginia Space Grant Consortium, and
the NSF grant AST-0807945.
S.R.M. acknowledges funding from NSF grants AST-0307851 and AST-0807945,
and NASA/JPL contract 1228235.
The National Radio Astronomy Observatory is operated by Associated Universities, Inc., 
under cooperative agreement with the National Science Foundation.

\appendix

\section{Details of the GBT Data Reduction}
\label{sec:gbtred}

\subsection{Baseline Removal and The Standing Wave}
\label{subsec:standingwave}

During the GBT observing we became aware of a sinusoidal pattern in velocity in the YY polarization
that made it difficult to use the standard polynomial baseline fitting routines in GBTIDL.  The sinusoidal
pattern is a known standing wave with a period of $\sim$1.5 MHz that arises from a total pathlength of $\sim$200 m and
a double reflection ($\pm\lambda$/8 focus shifts changing the phase by 180\degr).  The standing wave is highly
linearly polarized and only appears in the YY polarization \citep{Fisher03}.

\begin{figure*}[ht!]
\includegraphics[angle=0,scale=0.44]{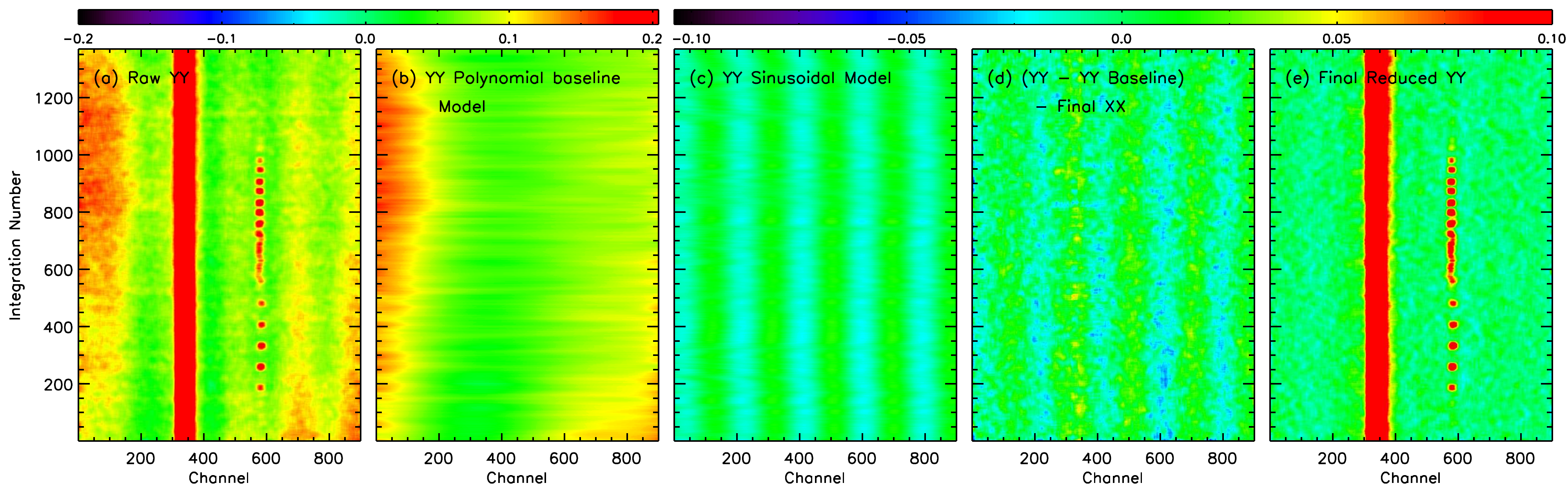}
\caption{YY polarization data for the GBT brick centered at ($l$,$b$)=(103.0\degr,$-40.0$\degr) showing our
GBT baseline reduction steps and the sinusoidal pattern.  The three-dimensional datacube
(position, position, velocity) is shown in a two-dimensional diagram by plotting the two spatial dimensions
on one axis as the sequential ``integration number''. ({\em a})  The raw YY data; ({\em b}) the YY-fitted 5th-order
polynomial baseline;  ({\em c}) the YY-fitted sinusoidal baseline model;  ({\em d}) the polynomial-baseline-subtracted
YY data minus the final XX reduced data (to remove real emission lines) showing the sinusoidal pattern in the data;
({\em e}) the final reduced YY data
with both the polynomial baseline and the sinusoidal pattern removed -- very little residual baseline is apparent.
Each diagram has been smoothed with a 20$\times$20 boxcar filter to make the
fainter features more visible. The units of the color scale are K.  The local Galactic zero-velocity gas and intermediate-velocity gas
is indicated by the red vertical stripe at Channel$\approx$350 and the Magellanic Stream is the vertical streak at
Channel$\approx$580.  The MS appears clumpy because the integration pattern scans across the MS and the local velocity
emission is continuous because it is pervasive.}
\label{fig_gbtreduction}
\end{figure*}

As described in \S \ref{subsec:gbtred}, we used our own special-purpose IDL routines to perform the baseline fitting
and removal.  The baseline fitting for the XX polarization was fairly straightforward, with a 5th-order polynomial
being fitted to the 21-cm spectrum (exluding the Galactic emission) with iterative emission line rejection.
Removing the sinusoidal pattern from the YY polarization was difficult because it was below the noise level of an individual
spectrum ($\sim$0.12 K),
and therefore multiple steps were required.  First, a 5th-order polynomial was iteratively fitted to the
spectrum excluding Galactic emission and the emission lines previously excluded for the XX polarization.
Next, a cosine was fitted to the spectrum with the initial polynomial baseline subtracted.  The amplitude and
wavelength were held fixed ($A$ = 0.02 K, $\lambda$ = 305 \kms [191 binned channels])
and only the phase was allowed to vary (with initial guess $\phi$ = 192 \kms [120 binned channels]).  Emission
lines were similarly excluded from the fit.  Finally, a polynomial$+$cosine function was fitted to the spectrum
using the previously derived parameters as initial guess, and allowing all parameters to vary.  The cosine
parameters were limited to $0.0 < A < 0.04$ K, $280 < \lambda <  330$ \kmse, and $0.0 < \phi < 400$ \kmse.
Galactic emission and emission lines previously excluded for the XX polarization and the initial YY cosine
fit were excluded during this final fitting process.  For some of the bricks the phase of the sinusoidal pattern
shifted slowly with time, but for others it stayed fairly constant.
We experimented with many variants of this method and parameter constraints until we converged on the procedure outlined above,
which is adequate for our purposes.

Figure \ref{fig_gbtreduction} shows an example of the YY data for one brick.  The three-dimensional datacube
(position, position, velocity) is shown in a two-dimensional diagram by collapsing the two spatial dimensions
on one axis as the sequential ``integration number''.  This makes it easier to visualize the overall features
in the data.  Each diagram has been smoothed with a 20$\times$20 boxcar filter to make the
fainter features more visible.

Our baseline removal process is not perfect and there are some wiggles at the mK level.  This is
probably due to the effects of incomplete rejection of channels contaminated by wings of emission lines slightly
skewing the baseline fits.  However, these baseline ripples do not affect our results for the MS.

\subsection{The GBT \hi Ghost}
\label{subsec:gbtghost}

\begin{figure}[t]
\begin{center}$
\begin{array}{cc}
\epsscale{1.00}
\includegraphics[angle=0,scale=0.40]{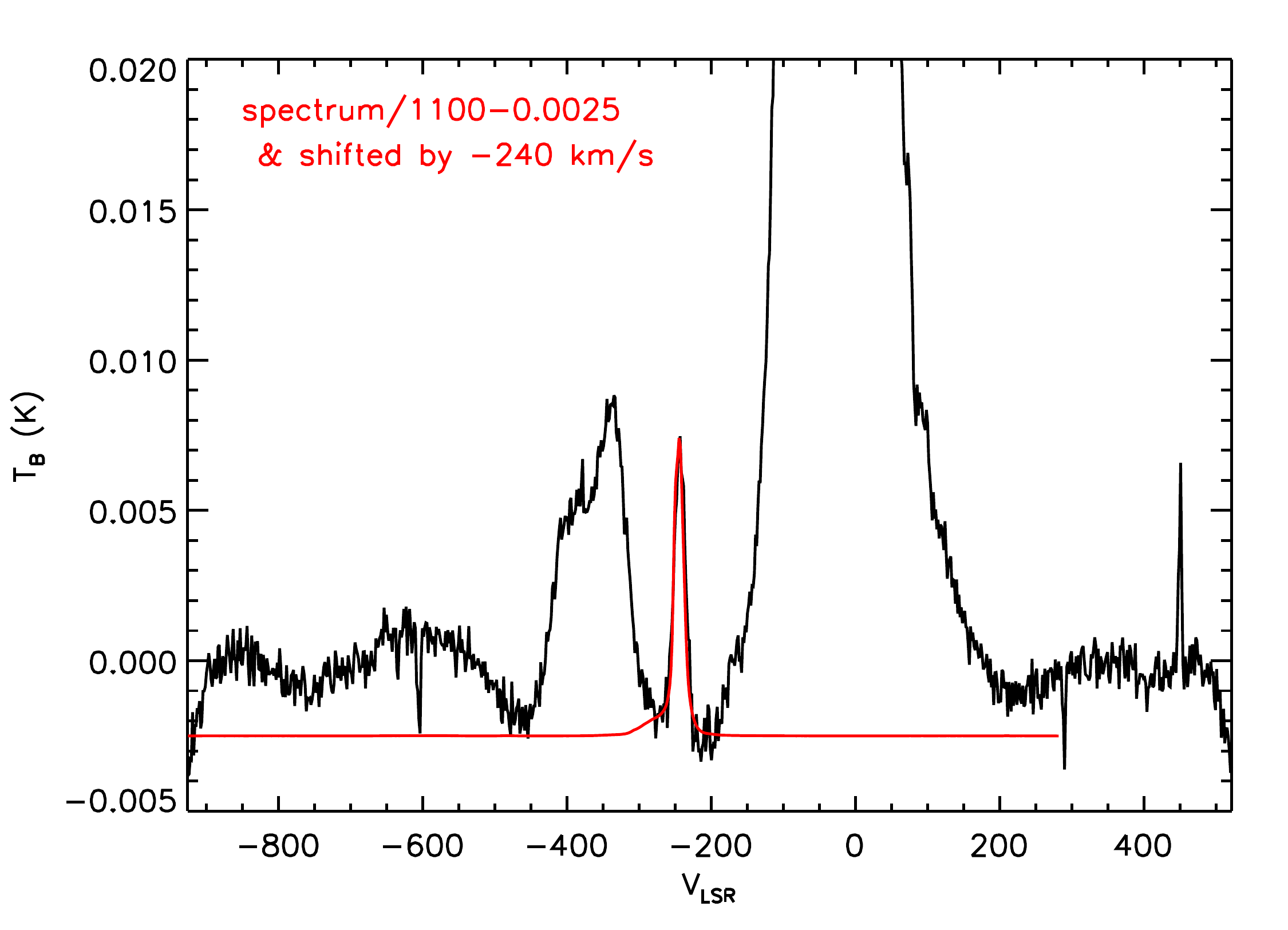} %
\includegraphics[angle=0,scale=0.38]{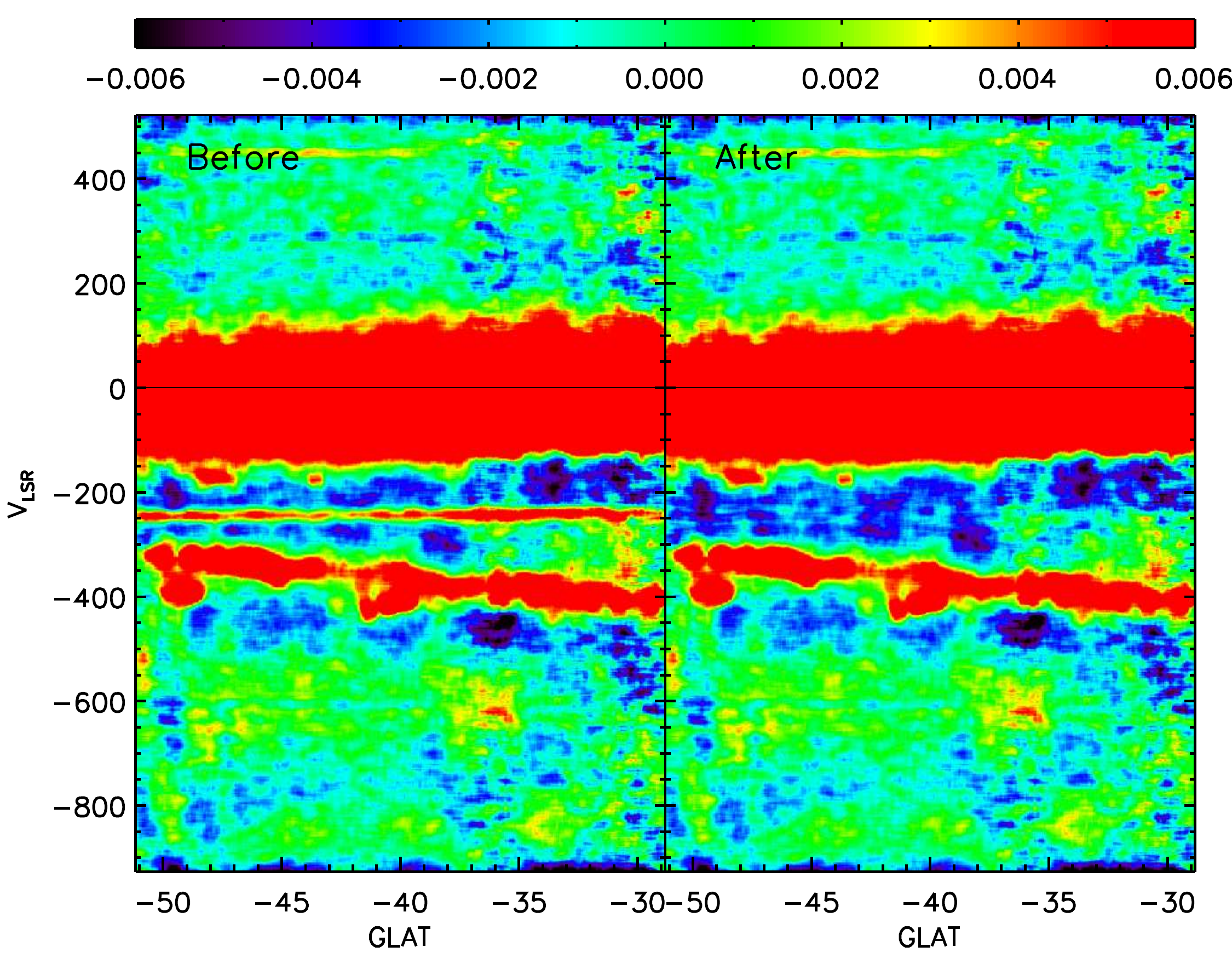}
\end{array}$
\end{center}
\caption{({\em Left}) Average spectrum of our entire GBT datacube.  The $-243.8$ \kms spurious emission line is clearly visible.  A
version of the average spectrum scaled by $1/1100$ and shifted by $-240$ \kms is overplotted in red and is an exact
match to the mysterious line.  This implies that the $-243.8$ \kms line is in fact a ``ghost'' spectrum in the
GBT data.  Some baseline ripples are still apparent in our data at the mK level, as are a few missed RFI lines.
({\em Right}) Intensities of the GBT data, averaged in Galactic longitude, in units of K.  The left panel shows the data before removal
of the $-243.8$ \kms ``ghost'' emission line, and the right panel shows the data after the ghost line
is removed by subtraction of a scaled and shifted version of the datacube from itself.  The unsmoothed datacube was used for the
figure and the final \vlsre--$b$ images were boxcar smoothed with a 20$\times$20 filter to enhance fainter features.}
\label{fig_gbt_ghost}
\end{figure}

There is a persistent, but weak, emission line permeating our entire GBT datacube (Fig.\ \ref{fig_gbt_ghost} left and Fig.\ 
\ref{fig_gbt_ghost} right).  This line does {\em not} appear in the S08 Arecibo datacube
or in the BT04 Westerbork datacube that both overlap portions of our survey and are both more sensitive than our data.  Therefore,
it is clear that this is not a real emission line, but a spurious artifact unique to the GBT data.   The line has a
Gaussian shape of $A$=0.004--0.013 K, \vlsr=$-243.8$ \kmse, and $\sigma_v$$\approx7.7$ \kmse, and appears in both
polarizations (with equal strength) as well as in the pre-baseline-subtracted, frequency-switched data (both
positive and negative images).  The suspect line maintains a constant velocity and line-width across our entire datacube.
The line is not radio frequency interference (RFI) because it is too wide ($\sim$200 unbinned channels) and because it
has a constant \vlsr (RFI should have a constant velocity in the observer rest frame).  The Gaussian shape of the
line suggests that it originates from an astronomical source.
However, it is unlikely that this is due to the stray radiation from forward spillover of the GBT secondary \citep{Lockman05}
because of the stability of the line (especially in \vlsr and $\sigma_v$) over some 200 deg$^2$ as well as the large negative
velocity.

The only astronomical \hi emission line that is so stable over such a large area is the local, Galactic zero-velocity emission.
In fact, the line-shape of the $-243.8$ \kms emission is a very close match to that of
the zero-velocity emission line.  The average spectrum of our entire GBT datacube is shown in Figure \ref{fig_gbt_ghost} left
with the $-243.8$ \kms emission line clearly revealed.  We have overplotted a version of the average spectrum scaled
by $1/1100$ and shifted by $-240$ \kms that {\em exactly} matches the mysterious emission line.  This suggests that the
$-243.8$ \kms emission line is in fact a ``ghost'' spectrum in the GBT data.  We attempted to remove the ghost by subtracting
a scaled and shifted version of the datacube from itself.  The ``before'' and ``after'' images of the GBT data (averaged in
Galactic longitude) are shown in Figure \ref{fig_gbt_ghost} right; the comparison indicates that the removal of the
line was successful, leaving no significant residual behind.

We propose that the ghost in our GBT data can be explained as a sideband with constant offset frequency on a local
oscillator used in the frequency conversion process that was not filtered out properly. This would cause a reproduction of
the input spectrum at a much lower amplitude and at the offset frequency, in this case $\sim$1.14 MHz, corresponding to
240 \kms at 21 cm.   It is possible that this ghost is present in other GBT \hi data and we suggest that other GBT observers
watch out for it.


\end{document}